\documentclass[a4paper,12pt]{article}

\usepackage[fleqn]{amsmath}
\usepackage{amsthm}
\usepackage[fleqn]{calculation}
\usepackage{txfonts}

\usepackage{tikz}
\usetikzlibrary{arrows.meta}
\usetikzlibrary{automata}
\usetikzlibrary{calc}
\newcommand{\scalefactor}{0.8}
\newcommand{\contentssize}{\Large}
\newcommand{\indexsize}{\normalsize}
\newcommand{\bracedescsize}{\footnotesize}
\newcommand{\snaptapedist}{5}

\newcommand{\Halmos}{$\Box$}

\newcommand{\sblank}{\texttt{\char32}}
\newcommand{\emptystring}{\varepsilon}

\newcommand{\funtype}[1]{\longrightarrow^{\textrm{#1}}}
\newcommand{\tfuntype}{\funtype{t}} 

\theoremstyle{definition}
\newtheorem{definitionThm}{Definition}

\newtheorem{remarkThm}[definitionThm]{Remark}

\newtheorem{lemmaThm}[definitionThm]{Lemma}

\newtheorem{proofThm}[definitionThm]{Proof}

\newcommand{\qstart}{{q_{\text{\textrm{start}}}}}
\newcommand{\qacc}{{q_{\text{\textrm{acc}}}}}
\newcommand{\qrej}{{q_{\text{\textrm{rej}}}}}

\newcommand{\dirleft}{\text{\textsf{left}}}
\newcommand{\dirright}{\text{\textsf{right}}}

\newcommand{\lparen}{\mathrm{LP}}
\newcommand{\rparen}{\mathrm{RP}}
\newcommand{\marksym}{\mathrm{X}}

\newcommand{\coqoptiontype}[1]{\mathcal{O}(#1)}
\newcommand{\coqsome}[1]{\lfloor#1\rfloor}
\newcommand{\coqnone}{\emptyset}

\newcommand{\coqconst}[1]{\mathsf{#1}}
\newcommand{\coqvar}[1]{\mathit{#1}}

\newcommand{\coqlab}[1]{\coqvar{#1}}
\newcommand{\coqTMtype}[2]{\coqconst{TM}_{#1}^{#2}}
\newcommand{\coqTMtypeSigmaOne}{\coqTMtype{\Sigma}{1}}
\newcommand{\coqTM}[2]{\coqvar{#1}^{(#2)}}

\newcommand{\coqswitch}{\coqconst{Switch}}
\newcommand{\coqmatchwith}[1]{\coqconst{match}\ {#1}\ \coqconst{with}}
\newcommand{\coqwhile}{\coqconst{While}}
\newcommand{\coqrelabel}{\coqconst{Relabel}}
\newcommand{\coqreturnopr}{\coqconst{Return}}
\newcommand{\coqreturn}[2]{\coqreturnopr_{#1}{#2}}

\newcommand{\coqread}[1]{\coqconst{Read}^{(#1)}}
\newcommand{\coqwrite}[1]{\coqconst{Write}^{(2)}\ {#1}}
\newcommand{\coqmove}[1]{\coqconst{Move}^{(2)}\ \mathrm{#1}}
\newcommand{\coqnop}{\coqconst{Nop}^{(1)}}

\title{
How to Verify a Turing Machine with Dafny}
\author{
Edgar F. A. Lederer\\
\mbox{}\\
Retired Lecturer from the\\
University of Applied Sciences and Arts\\
Northwestern Switzerland}
\date{}

\begin{document}

\maketitle

\begin{abstract}
This paper describes the formal verification of two Turing machines using the program verifier Dafny.
Both machines are deciders, so we prove total correctness.
They are typical first examples of Turing machines used in any course of Theoretical Computer Science;
in fact, the second machine is literally taken from a relevant textbook.
Usually, the correctness of such machines is made plausible by some informal explanations of their basic ideas, augmented with a few sample executions, but neither by rigorous mathematical nor mechanized formal proof.
No wonder:
The invariants (and variants) required for such proofs are big artifacts, peppered with overpowering technical details.
Finding and checking these artifacts without mechanical support is practically impossible, and such support is only available since recent times.
But nowadays, just because of these technicalities, with such subjects under proof a program verifier can really show off and demonstrate its capabilities.
\end{abstract}

\section{Introduction}

Are the Turing machines I used in my lectures ---and my examinations--- in Theoretical Computer Science \emph{really} doing what they are supposed to do?
Turing machines are quite fragile artifacts, so the answer to this quesion is not at all simple, let alone trivial.

To calm my conscience, I at least always tested my machines, using a Turing machine simulation infrastructure I have written in Haskell.
This was a healthy idea:
whenever I started developing a new machine, this testing uncovered errors in my initial designs.
Finally, I tested the machine on \emph{all} possible input words up to a certain length.
No errors were found anymore, and since I knew the internals of my machine and thus expected that it would not suddenly behave differently on input words longer than the longest used for testing, I started to \emph{believe} that my machine was indeed correct.

But how much better would it be to really \emph{know} that the machines are correct rather than just believing it?
Formal mechanical proof using a state-of-the-art program verifier could give a definite answer --- at least, sorry for the caveat, when \emph{knowing} the correctness of the verifier.
Nevertheless, even when just believing in its correctness, formal proof increases believing in the correctness of the subject under proof.
And definitely, it increases \emph{understanding} of the subject.

Over recent years I gained experience with the program verifier Dafny \cite{Leino:2023}.
Based on this experience I expected Dafny to be a suitable tool for my verification project.
To the best of my knowledge only a few researchers have worked on the formal verification of Turing machines, none of them using Dafny (Section~\ref{Section:RelatedWork}).
So my plan was ---besides the verification of the Turing machines themselves--- to demonstrate the suitability of Dafny for their verification.

Despite my trust in Dafny I suspected a verification of any Turing machine typical for teaching to be a challenge.
I started out with the verification of a Turing machine of my own, which decides the language of correctly nested and juxtaposed parentheses.
And indeed, this was challenging;
Section~\ref{Section:FirstExampleMachine} describes the development.
Then I was curious whether the techniques I had developed for this Turing machine could also be applied to another machine.
For this, I chose a machine from the textbook of Sipser \cite{Sipser:2006}, which decides the language of all words over a unary alphabet whose length is a power of two.
And indeed the techniques developed for the first machine were applicable to this second, but the development was quite different and even more challenging than the first;
Section~\ref{Section:SecondExampleMachine} describes the development.
Both developments can be downloaded from arXiv as ancillary files to this paper.

In the end, the two machines \emph{are} verified, and I like to conclude that Dafny \emph{is} suitable for their verification (Section~\ref{Section:Conclusions}).

\section{The Turing Machine Model}

There are many variants of Turing machine models, all of them equivalent in computational power.
Here we use the variant as defined by Sipser in his textbook \cite[Chapter~3]{Sipser:2006}, which we informally describe in the following.

\paragraph{Task}

The basic task of a Turing machine is to determine whether a given word over some input alphabet $\Sigma$ has, or does not have, a particular property.
If the Turing machine halts after a finite amount of time and outputs `accept', the input word has the considered property;
if the machine halts and outputs `reject', the input word does not have the property.
Sometimes, however, a Turing machine might never halt and thus might never produce a decision concerning the given word and the considered property.

\paragraph{Machine}

The basic ingredients of a Turing machine are a \emph{control unit}, a \emph{tape}, and a \emph{read-write-head}.
\begin{description}
\item[Control unit]
The control unit has a finite number of \emph{control states} or \emph{states} for short --- the set of these states is denoted by $Q$.
Some of these states are distinguished (and present in any Turing machine):
a \emph{start state} $\qstart \in Q$,
an \emph{accept state} $\qacc \in Q$, and
a \emph{reject state} $\qrej \in Q$.
Accept and reject state must be distinct.
(However, the start state might well be the accept state or the reject state.)
At any point in time, the control unit is in exactly one of these states.
\item[Tape]
The tape consists of a sequence of `squares'.
It has a beginning, but no end --- so it extends up to infinity.
At any point in time, each square contains a symbol taken from some \emph{tape alphabet} $\Gamma$.
We think of the tape as being realized in a horizontal fashion, with its beginning on the left side, and its extension up to infinity on the right side.
\item[Read-write-head]
At any point in time, the read-write-head, or simply \emph{head}, is positioned over exactly one square of the tape.
It can perform three actions in sequence:
First, it reads the symbol contained on that square.
Second, it overwrites this symbol with a new symbol.
(Note that `overwriting' entails the erasure of the old symbol, and thus (possible) loss of information.
Also note that the new symbol might be the same as the old one.
Engineers usually like to optimize this case by just doing nothing.)
Third, it moves exactly one square to the left or to the right --- to the left only if it is not at the beginning of the tape;
in this case it stays put.
\end{description}

\paragraph{Computation}

At any point in time, a Turing machine finds itself in an overall state, called a \emph{configuration} of the machine.
It is determined by three components:
(1) the current state of the control unit,
(2) the current tape contents, and
(3) the current position of the head.
A configuration is called an \emph{accepting (rejecting) configuration} if the state of control is the accept (reject) state, and it is called a \emph{halting configuration} if it is an accepting or rejecting configuration.

The computation of a Turing machine begins with a \emph{start configuration}, which is uniquely determined by the input word.
Then the machine goes step by step through a sequence of further configurations;
each configuration in this sequence is uniquely determined by its preceding configuration.
If the machine ever enters a halting configuration, computation stops.
Then, if this halting configuration is an accepting (a rejecting) one, the machine outputs `accept' (`reject').
If it never enters a halting configuration, it runs forever --- it \emph{loops}.

The start configuration is given as follows:
(1) The state of the control unit is the start state $\qstart$.
(2) The tape contents are the input word stored right at the beginning of the tape (the \emph{input alphabet} $\Sigma$ is a subset of the tape alphabet $\Gamma$), followed by an infinite sequence of \emph{blank symbols} $\sblank$.
A blank symbol is a distinguished (and present in any Turing machine) element of the tape alphabet.
However, it is not an element of the input alphabet.
So the end of the input word in the start configuration can be determined by the leftmost blank symbol.
(3) The position of the head is the beginning of the tape.

Given any non-halting configuration (also called \emph{continuing} configuration), the subsequent configuration is determined as follows:
First, the read-write-head reads the symbol at its current position.
Then, depending on the state of the control unit and the symbol just read, (1) the control unit enters a new state, (2) the head overwrites the old symbol with a new one, and (3) the head moves one position to the left (if possible) or to the right.
The new state, the symbol written, and the direction of move, are given by a total function%
\footnote{I use the arrow $\tfuntype$ to explicitly indicate a function to be total.}
\begin{align*}
\delta :
  Q \setminus \{\qacc, \qrej\} \times \Gamma \tfuntype
    Q \times \Gamma \times \{\dirleft, \dirright\}
\end{align*}
called the \emph{transition function}.
This function is at the heart of any Turing machine.
(Note that the source set of the function does not contain pairs with halting states;
clear --- halting configurations have no successor.)
We have now described the changes in the configuration.
But as important as what changes is what remains unchanged.
And unchanged are the complete tape contents except the overwritten symbol.
Observe that all changes in a configuration are very local concerning tape and head:
only the single symbol under the head is affected, and the head moves for (at most) one position.

\paragraph{How `infinite' is the tape?}

At any point in time, only a finite part of the tape can be filled with non-blank symbols.
The input is a word, by definition finite, and in each step at most one non-blank symbol may replace a blank symbol.
So practically it is not required for the tape to be infinite.
We could as well start the computation with a finite tape that just holds the input word, and whenever the machine runs out of tape, we glue further tape, filled with blank symbols, to the end of the tape already in use.
This would be similar to manually performing a computation on some sheets of paper.
Whenever the sheets are filled up with computations, we take a fresh ---blank--- further sheet.
We won't start a computation only after having organized an infinite amount of paper!

\section{The First Example Machine: Parentheses}
\label{Section:FirstExampleMachine}

The Turing machine we use as our first example is based on a machine I once developed (and tested) for some of my examinations in Theoretical Computer Science.
In fact, my original machine was unnecessarily more complicated than the machine presented here (in particular, it had a further control state);
only thanks to my attempts to verify the machines did I detect the simplifications.

\subsection{The Language}

This Turing machine decides the language $L$ over the input alphabet $\Sigma = \{(, )\}$ of correctly nested and justaposed parentheses.
This language can best be described by means of a context-free grammar
(with $S$ the only nonterminal symbol and thus the start symbol):
\begin{align*}
  S \rightarrow (S) \mid SS \mid \emptystring
\end{align*}
Examples of words in this language are $\emptystring$, `$()$', `$(())$', `$()()$', and `$((()())())$';
non-examples are `$)($', `$(()$' and `$())()$'.
Any context-free language can be recognized by a nondeterministic pushdown automaton.
However, for this particular language nondeterminism is not required.
Also, any context-free language can be decided by a Turing machine.
In fact, our example Turing machine simulates a deterministic pushdown automaton to decide $L$.

In Dafny we will represent any alphabet as datatype and any word over an alphabet as contents of an array of the corresponding datatype, where the length of the array equals the length of the represented word.
Here is the input alphabet as datatype:
\begin{verbatim}
datatype Sigma
  = LP // left parenthesis  '('
  | RP // right parenthesis ')'
\end{verbatim}

For the specification of our Turing machine (and a corresponding test oracle) we need a predicate that can be applied to an input array and that yields true if and only if its contents are a word of $L$.
One idea would be to formalize the context-free grammar given above together with derivations along this grammar.
However, I did not try this idea but followed a different approach.
In fact we see that a word is in $L$ iff it satisfies two conditions:
\begin{itemize}
\item (C1) the numbers of left and right parentheses in the word are the same, and
\item (C2) any prefix of the word contains no more right than left parentheses.
\end{itemize}
However, I must sadly admit that I have not formally proven these two conditions to be equivalent to our context-free grammar.

Here is a ghost function that yields the difference between the numbers of left and right parentheses in the first $i$ elements of array $a$:
\begin{verbatim}
ghost function leftMinusRightParen(a:array<Sigma>, i:int):int
  reads a
  requires 0 <= i <= a.Length
{
  if i == 0 then 0
  else
    match a[i-1]
    case LP => leftMinusRightParen(a, i-1) + 1
    case RP => leftMinusRightParen(a, i-1) - 1
}
\end{verbatim}
So the first condition translates into
\begin{verbatim}
leftMinusRightParen(a, a.Length) == 0
\end{verbatim}
And here is a ghost predicate%
\footnote{
Though not required by Dafny I always embrace quantifications in parentheses.
This gives a uniform appearance and a simple delineation of the scope of bound variables.
}
that yields true iff there are no more right than left parentheses in any prefix of the first $i$ elements of array $a$:
\begin{verbatim}
ghost predicate neverMoreRightThanLeftParen(a:array<Sigma>, i:int)
  reads a
  requires i <= a.Length
{
  (forall j :: 0 <= j <= i ==> leftMinusRightParen(a, j) >= 0)
}
\end{verbatim}
So the second condition translates into
\begin{verbatim}
neverMoreRightThanLeftParen(a, a.Length)
\end{verbatim}
Note that both the function and the predicate are formulated for an arbitrary prefix of the word, not just for the full word.
This generalization is later required for the corresponding invariants.
We already observe the necessity of the generalization of the function in the definition of the predicate.

Using the datatype declaration
\begin{verbatim}
datatype decision = accept | reject
\end{verbatim}
our Turing machine outputs \verb!accept! if and only if the word is in $L$.
What does the machine do if the word is not in $L$?
In fact, our machine is a \emph{decider}, so it terminates on all possible input words (which we will later prove, of course).
So if and only if the word is not in $L$, the machine outputs \verb!reject!.

Now we have all ingredients to specify our Turing machine:
\begin{verbatim}
method TMParentheses(a:array<Sigma>) returns (dec:decision)
  ensures dec == accept <==>
    neverMoreRightThanLeftParen(a, a.Length) &&
    leftMinusRightParen(a, a.Length) == 0
\end{verbatim}

We can also specify ---yes:\ and already implement and verify--- a \emph{test oracle} for our Turing machine.
Its specification is of course exactly the same (except for the name) as our machine.
Its implementation and verification follow the same ideas as our machine ---but are by far simpler--- and so provide a kind of high-level view onto the corresponding aspects.
This is very precious!

But do we really want to test a mechanism (and thus need a test oracle) when we can verify it?
Well, before starting with the verification, we want to be quite sure that our mechanism is indeed correct.
And this level of certainty can be obtained by means of testing.
Clearly the best test oracle we could employ is a verified one!

Here goes the oracle:
\begin{verbatim}
method OracleParentheses(a:array<Sigma>) returns (dec:decision)
  ensures dec == accept <==>
    neverMoreRightThanLeftParen(a, a.Length) &&
    leftMinusRightParen(a, a.Length) == 0
{
  var i := 0;
  var d := 0;
  while i < a.Length && 0 <= d
    invariant 0 <= i <= a.Length
    invariant d == leftMinusRightParen(a, i)
    invariant -1 <= d <= i
    invariant neverMoreRightThanLeftParen(a, i-1)
  {
    if a[i] == LP {
      d := d + 1;
    }
    else { assert a[i] == RP;
      d := d - 1;
    }
    i := i + 1;
  }
  dec := if d == 0 then accept else reject;
}
\end{verbatim}

Informally, the oracle works as follows:
It scans the input word from left to right.
In this process, it counts the number of parentheses, $+1$ for a left and $-1$ for a right parenthesis, and holds the result in a variable $d$.
So $d$ contains the difference of the number of left and right parentheses, which is expressed in the second invariant.

The process of counting continues as long as the input word has not yet been completely scanned \emph{and} the number of right parentheses does not surpass the number of left parentheses.
If there were more right than left parentheses, the difference $d$ would be $-1$, and the process would stop.
So even if more right parentheses would immediately follow, they would no longer be counted.
So $d$ can never fall below $-1$.
This is expressed in the (first conjunct) of the third invariant.

Though in a \emph{current} step (index $i$) the difference $d$ could be $-1$, in all \emph{previous} steps (all indices less than $i$) the difference was at least $0$, for otherwise the counting process would have stopped earlier.
This is expressed in the fourth invariant.

Upon termination (which itself is obvious) we conclude from the negation of the loop condition and the invariants that the input word has been completely scanned ($i = a.\mathit{Length}$; first disjunct) \emph{or} that the number of right parentheses surpasses the number of left parentheses by one ($d = -1$; second disjunct).
We distinguish three cases:
\begin{itemize}
\item Case 1: $d = 0$.
Since the second disjunct is not fulfilled, the first one must be.
So $i = a.\mathit{Length}$.
From the second invariant follows condition C1, and from the fourth invariant augmented with the second follows condition C2.
So the input word is accepted.
\item Case 2: $d > 0$.
Since the second disjunct is not fulfilled, the first one must be.
So $i = a.\mathit{Length}$.
From the second invariant follows violation of condition C1.
So the input word is rejected.
\item Case 3: $d = -1$.
Since the second disjunct \emph{is} fulfilled, we don't know anything about the first.
So $i \leq a.\mathit{Length}$.
So there exists some prefix of the input word that has more right than left parentheses, and thus condition C2 is violated.
So the input word is rejected.
\end{itemize}
We conclude:
The input word is accepted if and only if $d = 0$.

\subsection{The Machine}
\label{Subsection:MachineParentheses}

We have already presented the input alphabet $\Sigma = \{(, )\}$ of our Turing machine, together with its representation in Dafny as dataype \verb!Sigma!.
The tape alphabet of our machine is $\Gamma = \{\sblank, (, ), x, \$\}$.
We will call symbol $x$ `marking symbol' and symbol $\$$ `stack symbol'.
(As required, the tape alphabet contains a blank symbol $\sblank$, the blank symbol is not contained in the input alphabet, and the input alphabet is a subset of the tape alphabet.)
Here is our representation of the tape alphabet as datatype in Dafny:
\begin{verbatim}
datatype Gamma
  = B // blank symbol      ' '
  | L // left parenthesis  '('
  | R // right parenthesis ')'
  | X // marking symbol    'x'
  | S // stack symbol      '$'
\end{verbatim}

The tape contents will be represented as contents of an array of elements of type \verb!Gamma!.
Since \verb!Sigma! is not a subset (or subtype, respectively) of \verb!Gamma! we cannot directly store the input word on the tape.
Rather we represent input symbol \verb!LP! (\verb!RP!) of type \verb!Sigma! by tape symbol \verb!L! (\verb!R!) of type \verb!Gamma!.
The following function is used for this purpose:
\begin{verbatim}
function Sig2Gam(s:Sigma):Gamma
{
  match s
  case LP => L
  case RP => R
}
\end{verbatim}

The control states of our machine are
$Q = \{q_0, q_1, \ldots, q_5, \qacc, \qrej\}$,
the start state is $\qstart = q_0$.
As with the alphabets, we represent the states in Dafny as datatype:
\begin{verbatim}
datatype Q = q0 | q1 | q2 | q3 | q4 | q5 | q_acc | q_rej
\end{verbatim}

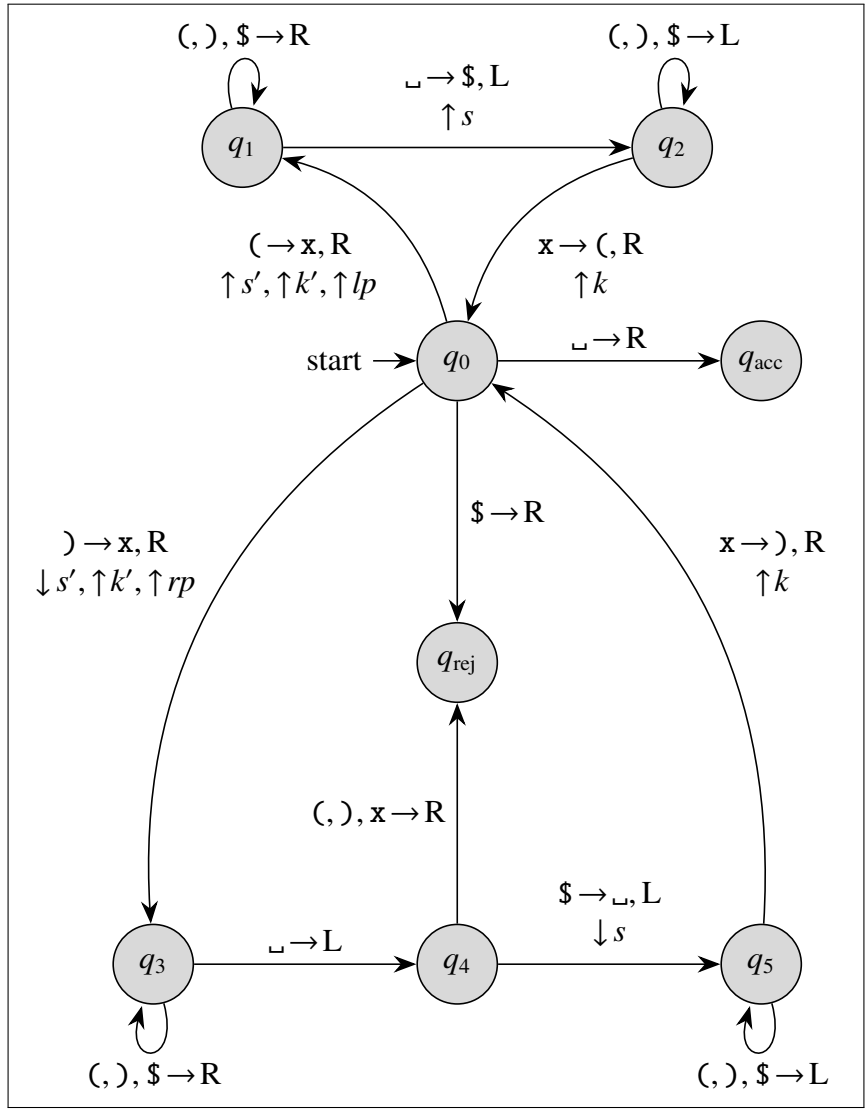
\begin{figure}[tb]
\centering
\framebox{
\begin{tikzpicture}[->,>={Stealth[scale=1.5]},shorten >=0pt,auto,
  node distance=4cm,semithick]

  \tikzstyle{every state}=[fill=black!15,draw]

  \node[initial,state] (q0)                       {$q_0$};
  \node[state]         (q1)   [above left of=q0]  {$q_1$};
  \node[state]         (q2)   [above right of=q0] {$q_2$};
  \node[state]         (qacc) [right of=q0]       {$\qacc$};
  \node[state]         (qrej) [below of=q0]       {$\qrej$};
  \node[state]         (q4)   [below of=qrej]     {$q_4$};
  \node[state]         (q3)   [left of=q4]        {$q_3$};
  \node[state]         (q5)   [right of=q4]       {$q_5$};

  \path
    (q0) edge [bend right] node
    {$\begin{matrix}
        \texttt{(}\!\rightarrow\!\texttt{x},\mathrm{R}\\
        \uparrow\!s', \uparrow\!k', \uparrow\!\mathit{lp}
      \end{matrix}$}
      (q1)
    (q0) edge node
    {$\sblank\!\rightarrow\!\mathrm{R}$}
      (qacc)
    (q0) edge node
    {$\texttt{\$}\!\rightarrow\!\mathrm{R}$}
      (qrej)
    (q0) edge [bend right] node [swap]
    {$\begin{matrix}
        \texttt{)}\!\rightarrow\!\texttt{x},\mathrm{R}\\
        \downarrow\!s', \uparrow\!k', \uparrow\!\mathit{rp}
      \end{matrix}$}
      (q3)
    (q1) edge [loop above] node
    {$\texttt{(},\texttt{)},\texttt{\$}\!\rightarrow\!\mathrm{R}$}
      ()
    (q1) edge node
    {$\begin{matrix}
        \sblank\!\rightarrow\!\texttt{\$},\mathrm{L}\\
        \uparrow\!s
      \end{matrix}$}
      (q2)
    (q2) edge [loop above] node
    {$\texttt{(},\texttt{)},\texttt{\$}\!\rightarrow\!\mathrm{L}$}
      ()
    (q2) edge [bend right] node
    {$\begin{matrix}
        \texttt{x}\!\rightarrow\!\texttt{(},\mathrm{R}\\
        \uparrow\!k
      \end{matrix}$}
      (q0)
    (q3) edge [loop below] node
    {$\texttt{(},\texttt{)},\texttt{\$}\!\rightarrow\!\mathrm{R}$}
      ()
    (q3) edge node
    {$\sblank\!\rightarrow\!\mathrm{L}$}
      (q4)
    (q4) edge node
    {$\texttt{(},\texttt{)},\texttt{x}\!\rightarrow\!\mathrm{R}$}
      (qrej)
    (q4) edge node
    {$\begin{matrix}
        \texttt{\$}\!\rightarrow\!\sblank,\mathrm{L}\\
        \downarrow\!s
      \end{matrix}$}
      (q5)
    (q5) edge [loop below] node
    {$\texttt{(},\texttt{)},\texttt{\$}\!\rightarrow\!\mathrm{L}$}
      ()
    (q5) edge [bend right] node [swap]
    {$\begin{matrix}
        \texttt{x}\!\rightarrow\!\texttt{)},\mathrm{R}\\
        \uparrow\!k
      \end{matrix}$}
      (q0);

\end{tikzpicture}
} 
\caption{First Turing machine --- Parentheses.
  \label{fig:state.transition.diagram.TMParentheses}}
\end{figure}

Figure~\ref{fig:state.transition.diagram.TMParentheses} shows the transition function of our machine (plus some additional information explained later) in the graphical form of a \emph{state transition diagram}.
By example, the labels on the transitions are read as follows:
\begin{itemize}
\item
Label `$(\ \rightarrow x, R$' on transition from control state $q_0$ to $q_1$ means:
If the control unit is in state $q_0$ and the head reads a $($, the control unit enters state $q_1$, the head overwrites the $($ with an $x$ and then moves to the right.
\item
Label `$(, ), \$ \rightarrow L$' on transition from control state $q_2$ to $q_2$ means:
If the control unit is in state $q_2$ and the head reads a $($, a $)$, or a $\$$, the control unit re-enters state $q_2$, the head overwrites the symbol just read with itself and then moves to the left (if possible).
\end{itemize}

Recall that the transition function is defined to be total.
However, not all transitions are explicitly shown in the diagram.
(Example: the transition leaving state $q_0$ with the head reading an $x$.)
Why?
The transitions not shown are never executed.
(Dafny will verify this fact, as explained later, and consequently, we will omit these transitions in our Dafny implementation --- they were just \emph{dead code}.)
However, for fulfilling the requirement for the transition function to be total we arbitrarily define all transitions not shown to go to the reject state and to move the head to the right.

\paragraph{Working of the machine --- informally}

Our machine scans the input word from left to right.
If it sees a left parenthesis, it overwrites the leftmost blank symbol on the tape with a stack symbol.
(This corresponds to a \emph{push} operation in a pushdown automaton.)
If it sees a right parenthesis \emph{and} there is a stack symbol on the tape, it overwrites the rightmost stack symbol on the tape with a blank symbol.
(This corresponds to a \emph{pop} operation in a pushdown automaton.)
If there is no stack symbol on the tape, the machine has seen exactly one more right parenthesis than left parentheses and thus enters the reject state (condition C2 violated).
However, if this situation never happens, the machine will completely scan the input word.
After this, it either sees a blank or a stack symbol.
If it sees a blank symbol, it has seen exactly as many left as right parentheses (condition C1 fulfilled), and it has never seen more right than left parentheses (condition C2 fulfilled).
Thus it enters the accept state.
If it sees a stack symbol, it has seen more left than right parentheses and thus enters the reject state (condition C1 violated).

\newlength{\fboxruleoriginal}
\setlength{\fboxruleoriginal}{\fboxrule}
\newlength{\headboxrule}
\setlength{\headboxrule}{2pt}

\newcommand{\tapealphabet}{$\sblank ( ) x \$$}
\newlength{\squarewidth}
\setlength{\squarewidth}{0.75cm}
\newcommand{\tapesquare}[1]
  {\framebox[\squarewidth]{\vphantom{\tapealphabet}#1}}

\newcommand{\headsquare}[1]{\setlength{\fboxrule}{\headboxrule}\tapesquare{#1}\setlength{\fboxrule}{\fboxruleoriginal}}

\newlength{\controlstatewidth}
\setlength{\controlstatewidth}{2\squarewidth}
\newcommand{\controlstate}[1]{\makebox[\controlstatewidth]{#1}}

\begin{figure}[tb]
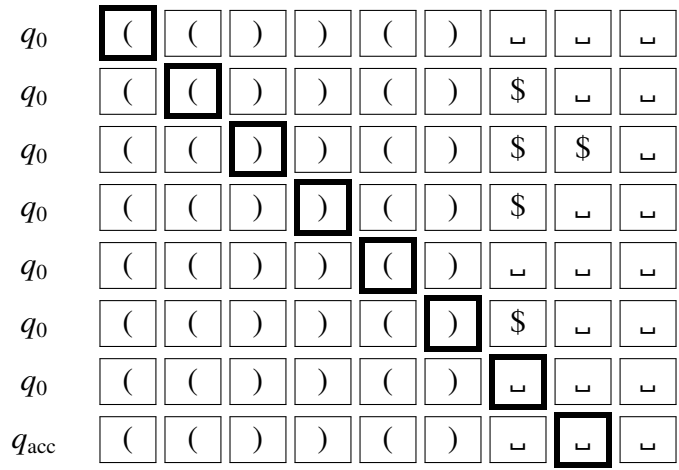

\begin{center}
  \controlstate{$q_0$}
    \headsquare{$($}
  \tapesquare{$($}
  \tapesquare{$)$}
  \tapesquare{$)$}
  \tapesquare{$($}
  \tapesquare{$)$}
  \tapesquare{$\sblank$}
  \tapesquare{$\sblank$}
  \tapesquare{$\sblank$}
    \mbox{}\\
  \controlstate{$q_0$}
  \tapesquare{$($}
    \headsquare{$($}
  \tapesquare{$)$}
  \tapesquare{$)$}
  \tapesquare{$($}
  \tapesquare{$)$}
  \tapesquare{$\$$}
  \tapesquare{$\sblank$}
  \tapesquare{$\sblank$}
    \mbox{}\\
  \controlstate{$q_0$}
  \tapesquare{$($}
  \tapesquare{$($}
    \headsquare{$)$}
  \tapesquare{$)$}
  \tapesquare{$($}
  \tapesquare{$)$}
  \tapesquare{$\$$}
  \tapesquare{$\$$}
  \tapesquare{$\sblank$}
    \mbox{}\\
  \controlstate{$q_0$}
  \tapesquare{$($}
  \tapesquare{$($}
  \tapesquare{$)$}
    \headsquare{$)$}
  \tapesquare{$($}
  \tapesquare{$)$}
  \tapesquare{$\$$}
  \tapesquare{$\sblank$}
  \tapesquare{$\sblank$}
    \mbox{}\\
  \controlstate{$q_0$}
  \tapesquare{$($}
  \tapesquare{$($}
  \tapesquare{$)$}
  \tapesquare{$)$}
    \headsquare{$($}
  \tapesquare{$)$}
  \tapesquare{$\sblank$}
  \tapesquare{$\sblank$}
  \tapesquare{$\sblank$}
    \mbox{}\\
  \controlstate{$q_0$}
  \tapesquare{$($}
  \tapesquare{$($}
  \tapesquare{$)$}
  \tapesquare{$)$}
  \tapesquare{$($}
    \headsquare{$)$}
  \tapesquare{$\$$}
  \tapesquare{$\sblank$}
  \tapesquare{$\sblank$}
    \mbox{}\\
  \controlstate{$q_0$}
  \tapesquare{$($}
  \tapesquare{$($}
  \tapesquare{$)$}
  \tapesquare{$)$}
  \tapesquare{$($}
  \tapesquare{$)$}
    \headsquare{$\sblank$}
  \tapesquare{$\sblank$}
  \tapesquare{$\sblank$}
    \mbox{}\\
  \controlstate{$\qacc$}
  \tapesquare{$($}
  \tapesquare{$($}
  \tapesquare{$)$}
  \tapesquare{$)$}
  \tapesquare{$($}
  \tapesquare{$)$}
  \tapesquare{$\sblank$}
    \headsquare{$\sblank$}
  \tapesquare{$\sblank$}
    \mbox{}
\end{center}
\caption{High-level view onto the computation.
  \label{fig:computation.TMParentheses.high-level.view}}
\end{figure}

Figure~\ref{fig:computation.TMParentheses.high-level.view} shows snapshots taken from the machine when it runs on (and accepts) input word `$(())()$'.
Any snapshot shows a configuration;
the head position is indicated by a thick frame.
All snapshots are taken when the machine is in control state $q_0$, except of the last one, when it is in the halting and accepting state $\qacc$.

Okay, but that was just a high-level view onto the computation --- after all, the parentheses are located somewhere on the left, and the stack symbols somewhere on the right part of the tape, but the head cannot jump over arbitrary distances.
Here is a lower-level view:
When the head sees a parenthesis, it replaces it with an $x$, the marking symbol.
The information whether the machine handles a left or a right parenthesis is held in the control state.
The head moves on to the right, square by square, until it finds a blank symbol.
There it adds or removes (if possible) a stack symbol.
After that, the head moves back, square by square, to the marking symbol.
It re-replaces the marking symbol with the left or right parenthesis the marking symbol has replaced.
Then, in the same way, the machine continues with the next parenthesis (if any).

\begin{figure}[tb]
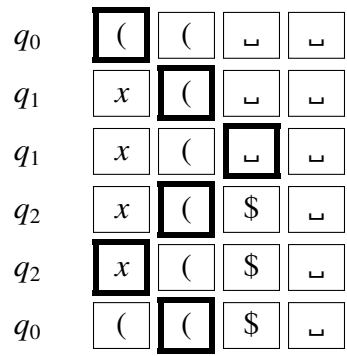

\begin{center}
  \controlstate{$q_0$}
    \headsquare{$($}
  \tapesquare{$($}
  \tapesquare{$\sblank$}
  \tapesquare{$\sblank$}
    \mbox{}\\
  \controlstate{$q_1$}
  \tapesquare{$x$}
    \headsquare{$($}
  \tapesquare{$\sblank$}
  \tapesquare{$\sblank$}
    \mbox{}\\
  \controlstate{$q_1$}
  \tapesquare{$x$}
  \tapesquare{$($}
    \headsquare{$\sblank$}
  \tapesquare{$\sblank$}
    \mbox{}\\
  \controlstate{$q_2$}
  \tapesquare{$x$}
    \headsquare{$($}
  \tapesquare{$\$$}
  \tapesquare{$\sblank$}
    \mbox{}\\
  \controlstate{$q_2$}
    \headsquare{$x$}
  \tapesquare{$($}
  \tapesquare{$\$$}
  \tapesquare{$\sblank$}
    \mbox{}\\
  \controlstate{$q_0$}
  \tapesquare{$($}
    \headsquare{$($}
  \tapesquare{$\$$}
  \tapesquare{$\sblank$}
    \mbox{}
\end{center}
\caption{Low-level view onto the computation.
  \label{fig:computation.TMParentheses.low-level.view}}
\end{figure}

Figure~\ref{fig:computation.TMParentheses.low-level.view} illustrates this process by showing \emph{all} configurations of the machine between and including the first two configurations with control state $q_0$ when running on input word `(('.

By the way:
Re-replacing the marking symbol with the original parenthesis is not necessary for deciding our language.
However, it entails no additional cost and seems to be elegant:
If the input word is accepted or rejected due to too many left parentheses, upon termination the tape contains the original input word;
if the input word is rejected due to a superfluous right parenthesis, the remaining marking symbol indicates the location of the problem.

\paragraph{How `finite' is the tape?}

Since our machine terminates on all input words (recall it is a decider), it only ever needs a finite amount of tape.
But how much?

Let $n$ denote the length of the input word.
If the input word contains nothing but left parentheses, the machine writes $n$ stack symbols to the right of the input word onto the tape.
So the length of the tape must be at least $2n$.
Furthermore, if the input word is empty ($n = 0$), the machine needs to read a blank symbol on the tape, so the length of the tape must be at least one.
Indeed, a tape length of $2n + 1$ always suffices;
Dafny will prove this fact.

A subtle point should be mentioned:
It may happen that the head moves one position beyond the rightmost square of the tape --- the head is then logically still located over a blank symbol, but technically over nothing.
(Indeed, this happens in our machine if and only if it accepts the empty word.)
However, this situation poses no problem as long as the head does not try to \emph{access} the tape, which means no problem as long as the machine stops.

\paragraph{Implementation}

Method \verb!TMParentheses! specified above yields the decision whether it accepts or rejects its input word, nothing else.
This is perfect concerning the actual task of the machine.
However, it would be very interesting to also know the tape contents upon termination.
We make this possible by putting the actual implementation of our machine into a more talkative second method \verb!TMParenImpl!;
the first method then calls this second one and ignores everything except of the decision.

Here we see (the major part of) the specification of the implementing method:
\begin{verbatim}
method TMParenImpl(a:array<Sigma>) returns
    (t:array<Gamma>, q:Q, ghost k:int, ghost s:int)
  ensures t.Length == 2 * a.Length + 1
  ensures q in {q_acc, q_rej}
  ensures q == q_acc <==>
    neverMoreRightThanLeftParen(a, a.Length) &&
    leftMinusRightParen(a, a.Length) == 0
\end{verbatim}
This method returns the tape contents, the final (halting) control state, and the contents of two ghost variables \verb!k! and \verb!s!.
The ghost variables will be used for the description of the tape contents during computation and upon termination;
two corresponding postconditions will be added later on.

Using this specification, we can implement and verify our first method:
\begin{verbatim}
method TMParentheses(a:array<Sigma>) returns (dec:decision)
  ensures dec == accept <==>
    neverMoreRightThanLeftParen(a, a.Length) &&
    leftMinusRightParen(a, a.Length) == 0
{
  var t, q;
  ghost var k, s;
  t, q, k, s := TMParenImpl(a);
  dec := if q == q_acc then accept else reject;
}
\end{verbatim}
With this, the first method \verb!TMParentheses! is finished;
we will talk only about the second one \verb!TMParenImpl! in the following.
And there we begin with completing the data structure for the configurations of the machine:
\begin{verbatim}
  t := new Gamma[2 * a.Length + 1];
  var p:nat;
\end{verbatim}
Variable \verb!p! contains the position of the head.
The variable for the control state is already present in the parameter list.

Now we load the start configuration.
The initial tape contents are the (transformed) input word on its left side and blank symbols everywhere else.
Using method
\begin{verbatim}
method copyInputToTape(a:array<Sigma>, t:array<Gamma>)
  modifies t
  requires t.Length == 2 * a.Length + 1
  ensures (forall i :: 0 <= i < a.Length ==> t[i] == Sig2Gam(a[i]))
  ensures (forall i :: a.Length <= i < t.Length ==> t[i] == B)
\end{verbatim}
we obtain the start configuration via
\begin{verbatim}
  copyInputToTape(a, t);
  p := 0;
  q := q0;
\end{verbatim}
Implementation and verification of method \verb!copyInputToTape! poses no further problems.

Realizing the computation of the machine is a straighforward translation of Figure~\ref{fig:state.transition.diagram.TMParentheses}:
\begin{verbatim}
  while q !in {q_acc, q_rej}
  {
    if q == q0 && t[p] == B {
      p := p + 1; q := q_acc;
    } else
    ...
    if q == q0 && t[p] == L {
      t[p] := X; p := p + 1; q := q1;
    } else
    ...
    if q == q2 && t[p] in {L, R, S} {
      if p > 0 {p := p - 1;}
    } else
    ...
    } else {
      // p := p + 1; q := q_rej;
      assert false;
    }
  }
\end{verbatim}

We see a single while-loop with a simple guard and a big conditional command as body.
The body is executed as long as the control state is not a halting one --- obvious.
The conditional command checks the control state and the symbol under the head.
Then, according to the transition function, it writes a new symbol onto the tape (or leaves it unaffected), changes the head position (if possible), and changes the control state (or leaves it unaffected).

Moving the head to the left is always guarded with the condition that it is not positioned at the beginning of the tape:
\begin{verbatim}
      if p > 0 {p := p - 1;}
\end{verbatim}
We don't make any efforts to analyze whether this condition is really required in any of the corresponding cases;
rather we always mechanically translate the rule that the head stays put at the beginning of the tape when it should go to the left.

Always using an `else' between the conditions and after the last condition ensures that \emph{all} combinations of control states and symbols that are not explicitly handled in one of the conditions are handled after the last `else'.
But these combinations correspond to exactly those transitions that are not shown in Figure~\ref{fig:state.transition.diagram.TMParentheses}, and these are ---as already mentioned--- never executed.
Dafny verifies this fact by proving the assertion `false' after the last `else'.
(Assertion `false' is never violated since control never reaches this point.)

\subsection{Verification}

\paragraph{Global and local invariants}

Since our implementation contains just a single loop, we need just a single invariant.
However, since this single loop expresses the complete functioning of our Turing machine, we can expect this single invariant to be quite large.
Large invariants are best expressed as conjunctions of smaller invariants;
we call all invariants of our loop (the conjuncts as well as the conjunctions) \emph{global} invariants.

What could be the structure of a global invariant?
When running the machine, we observe, for example, that the tape contains no marking symbol in states $q_0$ and $\qacc$ (and in $\qrej$ if it is reached via state $q_0$), but that it contains a single marking symbol in states $q_1$ up to $q_5$ (and in $\qrej$ if it is reached via state $q_4$).
We also observe, as another example, that the left parentheses are handled in states $q_1$ and $q_2$, and the right parentheses in states $q_3$, $q_4$, and $q_5$.
So each control state can put its own particular conditions onto the configuration of the machine;
in other words, each control state can have its own particular invariants.
We call these invariants \emph{local} invariants (meaning: local to the control state).

These observations suggest the following construction of a global invariant from local ones:
\begin{align*}
  (q = q_0 \implies I_0) \land
  \ldots \land
  (q = q_5 \implies I_5) \land \mbox{}\\
  (q = \qacc \implies I_{\text{\textrm{acc}}}) \land
  (q = \qrej \implies I_{\text{\textrm{rej}}})
\end{align*}
where $q$ denotes the current control state and $I_0$, {\ldots}, $I_5$, $I_{\text{\textrm{acc}}}$, $I_{\text{\textrm{rej}}}$ denote the local invariants.
(Note: $I_0$ is a local invariant; $q = q_0 \implies I_0$ is a global one.)

However, not all conditions on the configuration actually depend on the control state.
For example, the fact that the position of the head always lies between zero and the length of the tape is obviously independent of the control state;
it is expressed by the following global invariant:
\begin{verbatim}
    invariant 0 <= p <= t.Length
\end{verbatim}

\paragraph{Our proceeding}

We will develop our proof of correctness in five steps.
Finally at the end of each step the Dafny verification goes through.

In Step~1 we describe the tape contents, show how the number of stack symbols follows the number of parentheses, and relate the tape with the input word.
All this results in the first of two crucial global invariants, the `left-minus-right invariant'.
Dafny also shows the absence of runtime errors.
In Step~2 we consider the second of these two crucial global invariants, the `never-more invariant'.
In Step~3 we handle the postcondition and in Step~4 we prove termination.
Finally in Step~5 we add the postconditions promised above concerning the tape contents upon termination.

Practically, before showing the postcondition in Step~3 we must comment it out in the considered methods.
And before showing termination in Step~4 we must add a \verb!decreases *! clause to the loop and thus also to the method containing the loop and all methods directly or indirectly calling this method;
this stops Dafny from attempting to prove termination.

\paragraph{Step~1: The `left-minus-right invariant'}

In this first step we describe the tape contents in each control state, show that the left part of the tape is restored each time after handling a parenthesis, show that the number of stack symbols on the tape follows the number of parentheses already seen by the head, and finally show that the numbers of parentheses seen by the head corresponds to the numbers of parentheses in the input word.
All these considerations also enable Dafny to show the absence of runtime errors (all accesses to the tape are in range and the final false-assertion is unreachable).
Of course it would be very fine if we could prove these things separately (as is possible in the second example machine), but they seem to be so interwoven that this is not possible --- at least I did not find a more modular proof.

\subparagraph{Tape contents}

In any loop iteration the tape contents look as follows:
the left half of the tape contains the (transformed) input word, possibly with exactly one of the parentheses replaced by a marking symbol;
after that follow the stack symbols;
the remaining tape is filled with blank symbols.
The following two ghost predicates describe these conditions in detail, the first one for the case that there is no marking symbol on the tape, the second one for the case that there is exactly one:
\begin{verbatim}
ghost predicate tapeContentsWithoutX
    (a:array<Sigma>, t:array<Gamma>, s:nat)
  reads a, t
  requires t.Length == 2 * a.Length + 1
  requires 0 <= s <= a.Length
{
  (forall i :: 0 <= i < a.Length ==> t[i] == Sig2Gam(a[i])) &&
  (forall i :: a.Length <= i < a.Length + s ==> t[i] == S) &&
  (forall i :: a.Length + s <= i < t.Length ==> t[i] == B)
}
\end{verbatim}
\begin{verbatim}
ghost predicate tapeContentsWithXReplacing(paren:Sigma,
    a:array<Sigma>, t:array<Gamma>, k:nat, s:nat)
  reads a, t
  requires t.Length == 2 * a.Length + 1
  requires 0 <= k < a.Length
  requires 0 <= s <= a.Length
{
  a[k] == paren &&
  (forall i :: 0 <= i < a.Length && i != k ==>
      t[i] == Sig2Gam(a[i])) &&
  t[k] == X &&
  (forall i :: a.Length <= i < a.Length + s ==> t[i] == S) &&
  (forall i :: a.Length + s <= i < t.Length ==> t[i] == B)
}
\end{verbatim}
In both predicates parameters \verb!a! and \verb!t! denote the arrays containing the input word and the tape contents, respectively;
parameter \verb!s! denotes the number of stack symbols on the tape.
In the second predicate, parameter \verb!k! denotes the position of the marking symbol;
parameter \verb!paren! denotes the symbol (left or right parenthesis) that is replaced by the marking symbol --- this point will be explained later.

We use the ghost variables \verb!k! and \verb!s! declared in the parameter list of the implementing method to keep track of the number of parentheses already investigated (which equals the position of a possible marking symbol on the tape) and the number of stack symbols on the tape, respectively.
As global invariants we immediately get:
\begin{verbatim}
    invariant 0 <= s <= a.Length
    invariant 0 <= k <= a.Length
\end{verbatim}
Both variables are initialized to $0$;
\verb!k! is incremented when a marking symbol is re-replaced by its original parenthesis;
\verb!s! is incremented when a blank symbol is replaced by a stack symbol and decremented when a stack symbol is replaced by a blank symbol.
Here is an example from the loop body:
\begin{verbatim}
    ...
    if q == q2 && t[p] == X {
      t[p] := L; p := p + 1; q := q0;
      // ghost
      k := k + 1;
    } else
    ...
\end{verbatim}
All operations on ghost variables are shown in the state transition diagram in Figure~\ref{fig:state.transition.diagram.TMParentheses}, where we use notation $\uparrow\!x$ and $\downarrow\!x$ for incrementing and decrementing variable $x$, respectively.

Now we can formulate the invariants concerning the tape contents and can refine the already mentioned global invariants concerning \verb!p!, \verb!k!, and \verb!s! by taking the control state into account.
For example, in state $q_0$ we not only know that the head position \verb!p! is between zero and the tape length, but also that it equals the value of \verb!k!.
Here are the invariants:
\begin{verbatim}
    invariant q == q0 ==>
      0 <= s <= a.Length &&
      0 <= k <= a.Length &&
      tapeContentsWithoutX(a, t, s) &&
      k == p
    invariant q == q1 ==>
      0 <= s < a.Length &&
      0 <= k < a.Length &&
      tapeContentsWithXReplacing(LP, a, t, k, s) &&
      k < p <= a.Length + s
    invariant q == q2 ==>
      0 < s <= a.Length &&
      0 <= k < a.Length &&
      tapeContentsWithXReplacing(LP, a, t, k, s) &&
      k <= p < a.Length + s - 1
    invariant q == q3 ==>
      0 <= s < a.Length &&
      0 <= k < a.Length &&
      tapeContentsWithXReplacing(RP, a, t, k, s) &&
      k < p <= a.Length + s
    invariant q == q4 ==>
      0 <= s < a.Length &&
      0 <= k < a.Length &&
      tapeContentsWithXReplacing(RP, a, t, k, s) &&
      p == a.Length + s - 1
    invariant q == q5 ==>
      0 <= s < a.Length &&
      0 < k < a.Length &&
      tapeContentsWithXReplacing(RP, a, t, k, s) &&
      k <= p < a.Length + s
    invariant q == q_acc ==>
      0 == s &&
      k == a.Length &&
      tapeContentsWithoutX(a, t, 0) &&
      p == a.Length + 1
    invariant q == q_rej ==>
      (// via q0
       0 < s &&
       k == a.Length &&
       tapeContentsWithoutX(a, t, s) &&
       p == a.Length + 1)
         ||
      (// via q4
       0 == s &&
       0 <= k < a.Length &&
       tapeContentsWithXReplacing(RP, a, t, k, 0) &&
       p == a.Length)
\end{verbatim}
Note that the invariant for $\qrej$ takes into account that this state can be reached via two different states ($q_0$ and $q_4$), resulting in tape contents either without or with a marking symbol.

\subparagraph{Restoring the right (which could be a left --- pun intended) parenthesis}

When the Turing machine is in state $q_0$, then, according to the local invariant of $q_0$, the left half of the tape equals the (transformed) input word.
Now, when the head sees a left parenthesis (similar considerations apply to seeing a right parenthesis), it replaces this left parenthesis with a marking symbol, and the machine goes on into state $q_1$.
(The information that the machine now handles a left parenthesis has been moved from the tape to the control state.)
After possibly looping, the machine enters state $q_2$.
There, again after possibly looping, the head sees a marking symbol and replaces it with a left parenthesis, and the machine re-enters state $q_0$.
Since this marking symbol was the only one on the tape, and since the marking symbol is replaced by the same symbol (namely a left parenthesis) the marking symbol has replaced, the first half of the tape contents are restored to equal the (transformed) input word again.
This seems intuitively clear, but how can we convince Dafny?

As usual, an invariant helps.
But which one?
It is the condition $a[k] = \lparen$, which is a local invariant of states $q_1$ and $q_2$.
(In Dafny, we have already expressed this invariant by setting parameter \verb!paren! of the `$\marksym$-replacing predicate' in the local invariants for $q_1$ and $q_2$ to $\lparen$.)
\begin{description}
\item[Establishing the invariant]
Consider the Turing machine is in state $q_0$.
There, according to the local invariant of state $q_0$, the left half of the tape equals the transformed input word, and the head is at position $k$, thus $p = k$.
From that, when the head sees a left parenthesis, thus $t[p] = \lparen$ and thus $t[k] = \lparen$, we can conclude that the symbol in the input word at position $k$ must also be a left parenthesis, thus $a[k] = \lparen$.
(The left half of the tape and the transformed input word agree at all positions, thus in particular at position $k$.
Note that the invariant, which concerns the input word, is indirectly established by reading the tape --- the Turing machine cannot read the input word.)
\item[Maintaining the invariant]
When looping in state $q_1$, when going from $q_1$ to state $q_2$, and when looping in $q_2$, neither $k$ nor the contents of $a$ are modified.
(In fact, $a$ is not declared to be modifiable and thus simply cannot be modified.)
\item[Using the invariant]
Consider the Turing machine is in state $q_2$.
There, according to the local invariant of state $q_2$, the left half of the tape equals the transformed input word, but with the exception of position $k$, where it contains the marking symbol, thus $t[k] = \marksym$.
From that, in particular taking into account that there is only a single marking symbol on the tape, when the head (after possibly looping) sees a marking symbol, thus $t[p] = \marksym$, we can conclude that the head is at position $k$, thus $p = k$.

In order to restore the tape contents so that its left half again equals the transformed input word also at position $k$, the head must replace the marking symbol with the symbol in the input word at position $k$.
But this symbol, according to our new invariant $a[k] = \lparen$, is a left parenthesis.
And in fact, according to the state transition diagram, the head replaces the marking symbol with a left parenthesis.
\end{description}

\subparagraph{Counting parentheses with stack symbols}

Whether the machine enters the accept or reject state, or continues with the computation, depends on the existence of stack symbols on the tape.
And the number of stack symbols on the tape (and thus their existence) reflects the number of left and right parentheses seen by the head.
So we must establish a connection between these numbers of stack symbols and numbers of parentheses.
For this purpose we introduce four further ghost variables:
\verb!lp! and \verb!rp! contain the number of left and right parentheses, respectively, seen by the head from the beginning of computation up to its current point;
\verb!s'! and \verb!k'! contain difference and sum of those numbers of left and right parentheses, respectively.

All these variables are initialized with zero and incremented or decremented whenever the head sees a corresponding parenthesis in state $q_0$ and the machine goes on to state $q_1$ or $q_3$;
see the state transition diagram.
We obtain the following global invariants:
\begin{verbatim}
    invariant s' == lp - rp
    invariant k' == lp + rp
    invariant -1 <= s' <= k'
    invariant 0 <= k' <= a.Length
\end{verbatim}
Note that $s'$ is bounded from below by $-1$, though of course there can be much more right than left parentheses on the tape.
But if the difference is $-1$, computation and thus counting down stops (and the machine enters the reject state).

Variables \verb!s'! and \verb!s!, and \verb!k'! and \verb!k!, respectively, are thus strongly related.
In the end both variables \verb!s'! and \verb!s! refer to the difference between the numbers of left and right parentheses.
However, \verb!s'! is changed when the head sees a parenthesis, and \verb!s! when the head correspondingly writes or removes a stack symbol.
Likewise, in the end both variables \verb!k'! and \verb!k! refer to the total number of parentheses up to some point in computation.
However, \verb!k'! is incremented when the head replaces a parenthesis with a marking symbol, and \verb!k! when the head correspondingly re-replaces a marking symbol with a parenthesis.
So the actions on \verb!s! and \verb!k! just follow the corresponding actions on \verb!s'! and \verb!k'! later in time.
(All this holds but with one exception:
when the machine enters $\qrej$ via $q_4$, no actions on \verb!s! and \verb!k! will follow.)

The state transition diagram reveals the details, which we express as local invariants:
\begin{verbatim}
      s' == s     && k' == k       // for q0
      s' == s + 1 && k' == k + 1   // for q1
      s' == s     && k' == k + 1   // for q2
      s' == s - 1 && k' == k + 1   // for q3
      s' == s - 1 && k' == k + 1   // for q4
      s' == s     && k' == k + 1   // for q5
      s' == s     && k' == k       // for q_acc
      s' == s     && k' == k       // for q_rej via q0
      s' == s - 1 && k' == k + 1   // for q_rej via q4
\end{verbatim}
(Technically, we can add these local invariants as conjuncts to the local invariants seen above.)

\subparagraph{The `left-minus-right invariant'}

In the end we want to show that the input word is accepted if and only if our two conditions C1 and C2 are both fulfilled.
These conditions concern the input word, not the tape.
But the machine works on the tape.
So we must relate the number of parentheses on the tape with the number of parentheses in the input word.
The following crucial global `left-minus-right invariant' does the job:
\begin{verbatim}
    invariant s' == leftMinusRightParen(a, k')
\end{verbatim}
Spelled out:
The difference between the numbers of left and right parentheses on the \emph{tape} seen by the head equals the difference between the numbers of left and right parentheses in that prefix of the \emph{input word} whose length is the total number of left and right parentheses seen by the head.

And we enjoy good news: Now verification goes through!

\paragraph{Step~2: The `never-more invariant'}

When the machine detects a right parenthesis that has no left parenthesis as matching counterpart, it stops computation and rejects the input word.
This also means:
as long as the machine runs, the number of right parentheses seen by the head never exceeds the number of left parentheses previously seen by it.
But the tape just mirrors the input word.
So we can also say:
after having completely handled $k$ (say) parentheses, the machine has checked that there are never more right than left parentheses in the prefix of length $k$ of the input word.
This is expressed in the following crucial global `never-more invariant':
\begin{verbatim}
    invariant neverMoreRightThanLeftParen(a, k)
\end{verbatim}
And again we can be lucky:
Dafny proves this invariant without further help from our side.
This is fine: but why ---intuitively--- does it hold?

First we remark that in predicate \verb!neverMoreRightThanLeftParen! the precondition $i \leq a.\mathrm{Length}$ is fulfilled because of global invariant $k \leq a.\mathrm{Length}$ ($k$ actual parameter for formal parameter $i$).

Now, for easier notation we abbreviate function
$\lambda i \rightarrow \mathtt{leftMinusRightParen}(a, i)$
as $\Delta_a$ and function
$\lambda k \rightarrow \mathtt{neverMoreRightThanLeftParen}(a, k)$
as $\Theta_a$ --- the array $a$ containing the input word being fixed.
Our invariant then just reads as $\Theta_a(k)$.
So we rewrite the definitions of these two functions as follows:
\begin{align*}
  &\Delta_a : \{ 0 \ldots a.\mathrm{Length} \} \rightarrow \mathbb{Z}\\
  &\Delta_a(0) = 0\\
  &\Delta_a(i) =
    \begin{cases}
      \Delta_a(i-1) + 1 & \text{if $i \geq 1 \land a[i-1] = \lparen$}\\
      \Delta_a(i-1) - 1 & \text{if $i \geq 1 \land a[i-1] = \rparen$}
    \end{cases}
\end{align*}
and
\begin{align*}
  &\Theta_a : \{ \ldots a.\mathrm{Length} \} \rightarrow \mathbb{B}\\
  &\Theta_a(k) = (\forall j \mid
    0 \leq j \leq k \implies \Delta_a(j) \geq 0)
\end{align*}

In Appendix~\ref{Appendix:FirstMachine} we state and prove an \emph{Accumulated Invariant Lemma}.
It concerns a loop in which ---among other things--- a loop counter, $k$ say, counts up starting from zero.
The lemma basically says that if a condition $I(k)$ is an invariant of the loop, then the condition
\begin{align*}
  I_\forall(k) = (\forall j \mid 0 \leq j \leq k \implies I(j))
\end{align*}
is an invariant of the loop too.
Intuitively, the information obtained in the individual loop iterations gets accumulated.

Based on already discussed invariants, we will show in a moment that the condition $\Delta_a(k) \geq 0$ is a global invariant of our loop.
Then, according to the idea behind the Accumulated Invariant Lemma, the considered condition
\begin{align*}
  \Theta_a(k) =
    (\forall j \mid 0 \leq j \leq k \implies \Delta_a(j) \geq 0)
\end{align*}
is an invariant of our loop too.
(Alas, the lemma cannot be literally applied, since the loop body of our machine does not exactly match the form of the loop body in the lemma;
for example, $k$ in our machine is only incremented from time to time, and not in every loop iteration.)

Now we show that $\Delta_a(k) \geq 0$ is a global invariant of our loop.
Before starting with this endeavor we ask Dafny for confirmation:
\begin{verbatim}
    invariant 0 <= leftMinusRightParen(a, k) //explanational
\end{verbatim}
Such \emph{explanational} predicates (in invariants or assertions) are not needed by Dafny, but help us in our understanding or confirm that we are on the right track.

Base for our proof is the global `left-minus-right invariant', now expressed as
\begin{align*}
  s' = \Delta_a(k')
\end{align*}
We consider the individual control states and use their local invariants:
\begin{itemize}
\item $q_0$: Local invariants are:
$s' = s, k' = k, s \geq 0$.
Thus:
\begin{calculation}[\implies]
  s' = \Delta_a(k')
\step{$s' = s, k' = k$}
  s = \Delta_a(k)
\step{$s \geq 0$}
  \Delta_a(k) \geq 0
\end{calculation}
\item $q_1$: Local invariants are:
$s' = s+1, k' = k+1, s \geq 0, a[k] = \lparen$.
Thus:
\begin{calculation}[\implies]
  s' = \Delta_a(k')
\step{$s' = s+1, k' = k+1$}
  s+1 = \Delta_a(k+1)
\step{Def.\ of $\Delta_a$ with $a[k] = \lparen$}
  s+1 = \Delta_a(k) + 1
\step{Arith.}
  s = \Delta_a(k)
\step{$s \geq 0$}
  \Delta_a(k) \geq 0
\end{calculation}
\item $q_2$: Local invariants are:
$s' = s, k' = k+1, s \geq 1, a[k] = \lparen$.
Thus:
\begin{calculation}[\implies]
  s' = \Delta_a(k')
\step{$s' = s, k' = k+1$}
  s = \Delta_a(k+1)
\step{Def.\ of $\Delta_a$ with $a[k] = \lparen$}
  s = \Delta_a(k) + 1
\step{$s \geq 1$}
  \Delta_a(k) \geq 0
\end{calculation}
\end{itemize}
The remaining control states are handled in the same way (Appendix~\ref{Appendix:FirstMachine}).
We conclude:
since $\Delta_a(k) \geq 0$ is a local invariant of \emph{all} control states, it is a global one too.

\paragraph{Step~3: The postcondition}

Now it is time to harvest the fruits of our efforts:
the proof of the postcondition.
For easier reference we repeat the postcondition here:
\begin{verbatim}
  ensures q == q_acc <==>
    neverMoreRightThanLeftParen(a, a.Length) &&
    leftMinusRightParen(a, a.Length) == 0
\end{verbatim}
Again, Dafny automatically finds a proof, and again, we want to obtain a better understanding.
In our shorthand notation the postcondition reads as follows:
\begin{align*}
  q = \qacc \iff
    \Theta_a(a.\mathrm{Length}) \land \Delta_a(a.\mathrm{Length}) = 0
\end{align*}
It is an equivalence.
We will show in a moment that both conditions $\Theta_a(a.\mathrm{Length})$ and $\Delta_a(a.\mathrm{Length}) = 0$ are fulfilled if $q$ equals $\qacc$ (which proves direction $\Longrightarrow$), and that one of the two conditions is violated if $q$ equals $\qrej$.
But after termination $q$ either equals $\qacc$ or $\qrej$ --- negation of the loop guard.
So if $q$ does not equal $\qacc$, it must equal $\qrej$, and thus one of the two conditions is violated (which contrapositively proves direction $\Longleftarrow$).
\begin{itemize}
\item $q = \qacc$.
With the local invariants $s' = s, k' = k, s = 0, k = a.\mathrm{Length}$ the global invariants $\Theta_a(k)$ and $s' = \Delta_a(k')$ specialize to $\Theta_a(a.\mathrm{Length})$ and $\Delta_a(a.\mathrm{Length}) = 0$.
So both conditions are fulfilled.
\item $q = \qrej$ via $q_0$.
With local invariants $s' = s, k' = k, s > 0, k = a.\mathrm{Length}$ the global invariants $\Theta_a(k)$ and $s' = \Delta_a(k')$ specialize to $\Theta_a(a.\mathrm{Length})$ and $\Delta_a(a.\mathrm{Length}) > 0$.
So the second condition is violated (the first one is fulfilled).
\item $q = \qrej$ via $q_4$.
With local invariants $s' = s-1, k' = k+1, s = 0, 0 \leq k < a.\mathrm{Length}$ the global invariant $s' = \Delta_a(k')$ specializes to $\Delta_a(k+1) = -1$;
also we have $1 \leq k+1 \leq a.\mathrm{Length}$, thus $0 \leq k+1 \leq a.\mathrm{Length}$.
Now we show that $\Theta_a(a.\mathrm{Length})$ is violated.
We calculate%
\footnote{
In hints given in calculational proofs I sometimes refer to the textbook of Gries and Schneider \cite{Gries/Schneider:1993}.
For example, (GS~9.13) refers to Law (9.13), Instantiation, in that book.
However, the corresponding steps should be easily understandable also without looking it up there.
}:
\begin{calculation}[\implies]
  \Theta_a(a.\mathrm{Length})
\step{Def.\ of $\Theta_a$}
  (\forall j \mid
    0 \leq j \leq a.\mathrm{Length} \implies \Delta_a(j) \geq 0)
\step{Instantiation with $j = k+1$ (GS 9.13)}
  0 \leq k+1 \leq a.\mathrm{Length} \implies \Delta_a(k+1) \geq 0
\step{Modus ponens with $0 \leq k+1 \leq a.\mathrm{Length}$}
  \Delta_a(k+1) \geq 0
\end{calculation}
However, we already found $\Delta_a(k+1) = -1$.
So we can read these implications contrapositively and conclude that $\Theta_a(a.\mathrm{Length})$, the first condition, is violated.
(The second condition, $\Delta_a(a.\mathrm{Length}) = 0$, remains unchecked for $k+1 < a.\mathrm{Length}$, and is violated with $\Delta_a(a.\mathrm{Length}) = -1$ for $k+1 = a.\mathrm{Length}$.)
\end{itemize}

With these considerations we conclude our proof of partial correctness of our Turing machine:
Whenever the machine halts, it yields the correct result.

\paragraph{Step~4: Termination}

But does it always halt?
Yes, and here is the proof!

The Turing machine scans its input parenthesis by parenthesis.
With each parenthesis completely handled, the value of variable \verb!k! is incremented by one.
So the value of expression `\verb!a.Length - k!' decreases with each parenthesis handled by one.
Global invariant $k \leq a.\mathrm{Length}$ tells us that the expression is bounded from below by zero.

However, handling a single parenthesis requires to visit several states of the control unit.
So the machine makes progress when it goes from state $q_0$ to state $q_1$ (or to $q_3$, respectively), and then further to state $q_2$.
Then going from state $q_2$ back to $q_0$ looks like a step backwards --- however, in this moment \verb!k! is incremented.
So we must use a lexicographic ordering, with $k$ the major and the state the minor criterion.
But how can we express this ordering on the states?
Using ghost function
\begin{verbatim}
ghost function orderQ(q:Q):int
{
  match q
  case q0 => 0
  case q1 => 1
  case q2 => 2
  case q3 => 1
  case q4 => 2
  case q5 => 3
  case q_acc => 1
  case q_rej => 3
}
\end{verbatim}
the value of expression `\verb!3 - orderQ(q)!' decreases with each progress from one state to the next (except of going back to $q_0$), and is bounded from below by zero.

However, the machine does not always go from a state to a different state, but sometimes goes from a state back to itself.
Then the machine makes progress just by moving the head.
If the head moves to the right (states $q_1$ and $q_3$), the value of expression `\verb!t.Length - p!' decreases, and if the head moves to the left (states $q_2$ and $q_5$), the value of expression `\verb!p!' decreases.
Global invariant $0 \leq p \leq t.\mathrm{Length}$ tells us that both expressions are bounded from below by zero.
When the machine leaves any of these selfies, the direction of the head movement generally changes (in our particular machine it always changes).
This again looks like a step backwards --- however, in this moment progress is made in the control state ($q_1$ to $q_2$ or $q_3$ to $q_4$) or the value of \verb!k! ($q_2$ or $q_5$ to $q_0$).
Again, we need a lexicographic ordering, now altogether threefold, with $k$ the major, the state the middle, and the position of the head the minor criterion.

All this can be expressed in Dafny as follows:
\begin{verbatim}
  decreases a.Length - k, 3 - orderQ(q),
    if q in {q1, q3} then t.Length - p else p
\end{verbatim}

This concludes our proof of termination, and together with our proof of partial correctness also our proof of total correctness.

\paragraph{Step~5: Tape contents upon termination}

Recall that we have introduced the more talkative method for the implementation of our machine in order to also know the tape contents upon termination, and that we have promised to add two further corresponding postconditions to this method.
Here they are:
\begin{verbatim}
  ensures q == q_acc ==>
    0 == s &&
    k == a.Length &&
    tapeContentsWithoutX(a, t, 0) &&
    0 == leftMinusRightParen(a, a.Length) &&
    a.Length % 2 == 0
  ensures q == q_rej ==>
    (// via q0
     0 < s <= a.Length &&
     0 < k == a.Length &&
     tapeContentsWithoutX(a, t, s) &&
     0 < leftMinusRightParen(a, a.Length))
       ||
    (// via q4
     0 == s < a.Length &&
     0 <= k < a.Length &&
     tapeContentsWithXReplacing(RP, a, t, k, 0) &&
     -1 == leftMinusRightParen(a, k + 1))
\end{verbatim}

These added postconditions are immediate consequences of what we have developed up to now, and Dafny proves them automatically.
However, a few words about the added postconditions concerning the length of the input word might be useful.
In the accept-case, this length must be even, and in the reject-case, it must be at least one.
It might be interesting to note that these postconditions don't need the mechanics of the machine to be established, rather they immediately follow from the already existing postconditions.
According to these original postconditions the input word of length zero is accepted.
So any rejected input word must have a length of at least one.
And any accepted input word ---again according to the original postconditions--- has the same number of left and right parentheses.
But if we have $n$ (say) apples and the same number of pears, we altogether have $2n$ fruits.

\section{The Second Example Machine: Sipser's $M_2$}
\label{Section:SecondExampleMachine}

The Turing machine we use as our second example is the Turing machine $M_2$ presented as Example $3.7$ in the textbook of Michael Sipser \cite[Chapter~3]{Sipser:2006}.

\subsection{The Language}

This Turing machine decides the language over the input alphabet $\Sigma = \{ 0 \}$ of all words whose length is a power of two ($\mathbb{N}$ includes $0$):
\begin{align*}
  L = \{\,0^n \mid n \in \mathbb{N} \land
                   (\exists k \in \mathbb{N} \mid n = 2^k)\,\}
    = \{ 0, 00, 0000, 00000000, \ldots \}
\end{align*}

We can naturally specify our Turing machine as follows:
\begin{verbatim}
method TMSipserM2(a:array<Sigma>) returns (dec:decision)
  ensures dec == accept <==> isPowerOf2(a.Length)
\end{verbatim}
where we use the following straightforward definitions:
\begin{verbatim}
datatype Sigma = Zero // zero '0'
datatype decision = accept | reject
ghost predicate isPowerOf2(n:nat)
{
  (exists k:nat :: n == power(2, k))
}
ghost function power(b:int, k:nat):int
{
  if k == 0 then 1 else b * power(b, k-1)
}
\end{verbatim}

But how can we decide whether a natural number is a power of two?
We use the following three quite obvious lemmas:
\begin{verbatim}
lemma isPowerOf2LemmaZero()
  ensures !isPowerOf2(0)
\end{verbatim}
stating that zero is not a power of two;
\begin{verbatim}
lemma isPowerOf2LemmaEven(n:nat)
  ensures n % 2 == 0 ==> (isPowerOf2(n) <==> isPowerOf2(n / 2))
\end{verbatim}
stating that an even natural number is a power of two iff half of it is a power of two;
\begin{verbatim}
lemma isPowerOf2LemmaOdd(n:nat)
  ensures isPowerOf2(1)
  ensures n % 2 != 0 ==> (isPowerOf2(n) ==> n == 1)
\end{verbatim}
stating that one is a power of two and that it is the only odd natural number that is a power of two.
Dafny proofs of these lemmas are given in Appendix~\ref{Appendix:SecondMachine}.

Now we devise the following algorithm:
If the input number is zero, it is rejected, according to the first lemma.
Otherwise, we compute a sequence of numbers, starting with the input number, and obtaining each further number in this sequence by halving its immediate predecessor --- as long as this predecessor was even.
According to the second lemma, all numbers in the sequence maintain the property of either or not being a power of two.
The sequence obviously ends, and obviously ends with an odd number.
If and only if this odd number equals one, the input number is accepted, according to the third lemma.

We will employ this algorithm in our Turing machine as well as in the following test oracle:
\begin{verbatim}
method OracleSipserM2(n:nat) returns (acc:bool)
  ensures acc <==> isPowerOf2(n)
{
  if n == 0 {
    isPowerOf2LemmaZero();
    acc := false;
  }
  else {
    var s:nat := n;
    while s % 2 == 0
      invariant s >= 1
      invariant isPowerOf2(s) <==> isPowerOf2(n)
      decreases s
    {
      isPowerOf2LemmaEven(s);
      s := s / 2;
    }
    isPowerOf2LemmaOdd(s);
    acc := s == 1;
  }
}
\end{verbatim}
(Since our input alphabet has just a single symbol, the input to the machine is completely determined by its length.
So, for simplicity, the oracle just takes a number as input, representing the input array of the Turing machine under test.
Also for simplicity, we just return the decision as a boolean.)

\subsection{The Machine}

The tape alphabet of our machine is $\Gamma = \{\sblank, 0, x\}$;
we will call symbol $x$ `cross symbol'.
Here are the datatype for the tape alphabet as well as the function for representing the input symbol \verb!Zero! of type \verb!Sigma! by the tape symbol \verb!Z! of type \verb!Gamma!:
\begin{verbatim}
datatype Gamma =
    B // blank symbol ' '
  | Z // zero         '0'
  | X // cross symbol 'x'

function Sig2Gam(s:Sigma):Gamma
{
  match s
  case Zero => Z
}
\end{verbatim}
(The matching is done in analogy to our first machine;
here it is clearly not required.)

The control states of our machine are
$Q = \{q_0, q_1, \ldots, q_4, \qacc, \qrej\}$,
the start state is $\qstart = q_0$.
Here in Dafny:
\begin{verbatim}
datatype Q = q0 | q1 | q2 | q3 | q4 | q_acc | q_rej
\end{verbatim}

\begin{figure}[tb]
\centering
\framebox{
\begin{tikzpicture}[scale=0.9,->,>={Stealth[scale=1.5]},
  shorten >=0pt,auto,semithick]

  \tikzstyle{every state}=[fill=black!15,draw]

  \path[step=1cm,gray,very thin] (-3,-7) grid +(16,14);

  \node[initial,state] (q0)   at +(0,0)   {$q_0$};
  \node[state]         (q1)   at +(5,0)   {$q_1$};
  \node[state]         (q2)   at +(10,0)  {$q_2$};
  \node[state]         (q3)   at +(10,-4) {$q_3$};
  \node[state]         (q4)   at +(7.5,4) {$q_4$};
  \node[state]         (qacc) at +(5,-4)  {$\qacc$};
  \node[state]         (qrej) at +(0,-4)  {$\qrej$};

  \path
    (q0) edge node
    {$\begin{matrix}
        \texttt{0}\!\rightarrow\!\sblank,\mathrm{R}\\
        \mathit{snapGlob} := t
      \end{matrix}$}
      (q1)
    (q0) edge node
    {$\sblank,\texttt{x}\!\rightarrow\!\mathrm{R}$}
      (qrej)
    (q1) edge node
    {$\texttt{0}\!\rightarrow\!\texttt{x},\mathrm{R}$}
      (q2)
    (q1) edge node
    {$\sblank\!\rightarrow\!\mathrm{R}$}
      (qacc)
    (q2) edge [bend left] node
    {$\texttt{0}\!\rightarrow\!\mathrm{R}$}
      (q3)
    (q2) edge node [swap]
    {$\sblank\!\rightarrow\!\mathrm{L}$}
      (q4)
    (q3) edge [bend left] node
    {$\texttt{0}\!\rightarrow\!\texttt{x},\mathrm{R}$}
      (q2)
    (q3) edge [bend left] node
    {$\sblank\!\rightarrow\!\mathrm{R}$}
      (qrej)
    (q4) edge node [swap]
    {$\begin{matrix}
        \sblank\!\rightarrow\!\mathrm{R}\\
        \mathit{snapGlob} := t
      \end{matrix}$}
      (q1)
    (q1) edge [loop above] node
    {$\texttt{x}\!\rightarrow\!\mathrm{R}$}
      ()
    (q2) edge [loop above] node
    {$\texttt{x}\!\rightarrow\!\mathrm{R}$}
      ()
    (q3) edge [loop below] node
    {$\texttt{x}\!\rightarrow\!\mathrm{R}$}
      ()
    (q4) edge [loop above] node
    {$\texttt{0},\texttt{x}\!\rightarrow\!\mathrm{L}$}
      ();

\end{tikzpicture}
} 
\caption{Second Turing machine --- Powers of two (Sipser's $M_2$).
  \label{fig:state.transition.diagram.TMSipserM2}}
\end{figure}
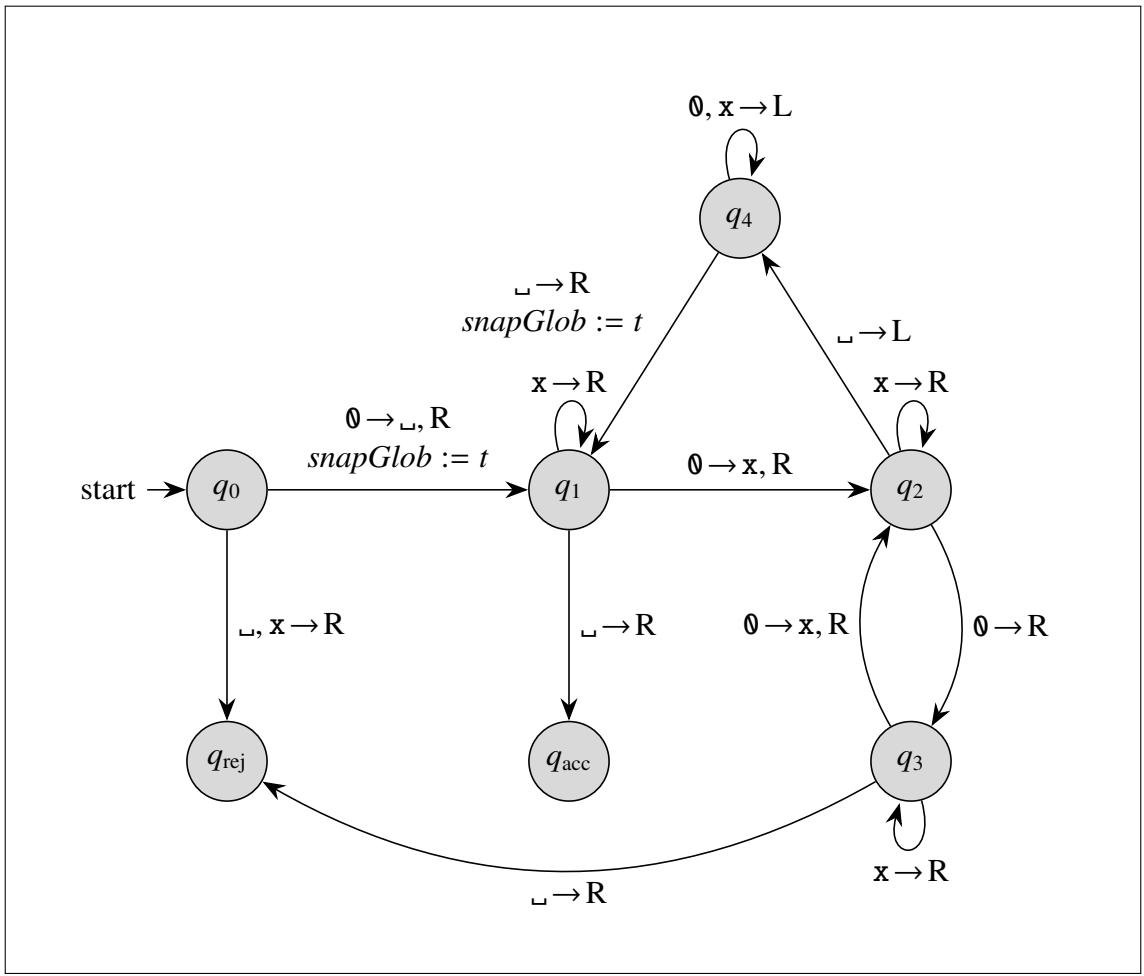

Figure~\ref{fig:state.transition.diagram.TMSipserM2} shows the state transition diagram of our machine.
There is a single transition that can never be executed (from $q_0$ when reading $x$);
for ensuring the totality of the transition function we arbitrarily let this transition go to $\qrej$ and move the head to the right.
However, the Dafny implementation has no code for this transition;
Dafny rather verifies that indeed this transition can never be executed.

\paragraph{Working of the machine --- informally}

First:
The machine needs to repeatedly perform integer divisions of the number of zeroes still found on the tape by two, simultaneously producing a quotient and a remainder.
The computation of the quotient is performed by crossing off each second zero, and the computation of the remainder by repeatedly changing between states $q_2$ and $q_3$, after starting with $q_1$.
(In a sense our Turing machine is more efficient than our test oracle:
The Turing machine computes quotient and remainder simultaneously.
The test oracle, however, performs each division (except of the last one) twice:
first in the loop guard to obtain the remainder (throwing the quotient away), and then in the loop body to obtain the quotient (throwing the remainder away).)

Second:
After performing an integer division resulting in an even number of zeroes, the machine needs to move its head back to the beginning of the tape.
In order to determine this beginning, the machine uses a standard trick:
It replaces the very first zero symbol on the tape, but no other, with a blank symbol ($q_0$ to $q_1$);
moving the head back (looping on $q_4$) stops when finding a blank symbol ($q_4$ to $q_1$).
Since the only blank symbol on the way backwards is the one initially written, moving backwards indeed stops at the beginning of the tape.

But carefully note:
That the zero symbol at the beginning of the tape is replaced by a blank symbol does not mean that it is crossed off;
the first blank symbol is considered to be just another representation of zero.

Now the working:
We distinguish three cases:
\begin{itemize}
\item
If there are no zeroes in the input, the machine enters the reject state, according to the first lemma ($q_0$ then $\qrej$).
\item
If there is a single zero in the input, the machine enters the accept state, according to the third lemma ($q_0$ then $q_1$ then $\qacc$).
\item
If there are at least two zeroes in the input, the machine repeatedly scans (the relevant prefix of) its tape from left to right.
In each such scan, it performs the integer division described above, not changing the truth or falsity of the `is-a-power-of-two predicate' of the number of zeroes on the tape, according to the second lemma.
We distinguish two cases:
\begin{itemize}
\item
If the number of zeroes in a scan was even (there were at least two zeroes; the scan ends up in $q_2$), the next scan is performed (after moving the head back to the beginning of the tape via looping on $q_4$).
For this next scan we distinguish two cases:
\begin{itemize}
\item
If the quotient is one, the scan ends up in the accept state, according to the third lemma (after reading only cross symbols via looping on $q_1$).
\item
If the quotient is at least two, the scan performs a further division by two.
\end{itemize}
\item
If the number of zeroes in a scan was odd (there were at least three zeroes; the scan ends up in $q_3$), the machine enters the reject state, according to the third lemma ($q_3$ then $\qrej$).
\end{itemize}
\end{itemize}

\paragraph{How `finite' is the tape?}

Let $n$ denote the length of the input word.
Scanning the input word from left to right stops when reading the blank symbol immediately following the input word.
Computation then either stops or the head moves back to the beginning of the tape.
So a length of $n+1$ for the tape is required as well as sufficient.

Note:
With this tape length, the head always moves one position beyond the end of the tape when computation stops.
But since computation stops, the head does no longer access the tape --- so there is no problem!

\paragraph{Implementation}

The basic ideas are completely analogous to our first Turing machine;
please also refer to the explanations given in the corresponding paragraph in Subsection~\ref{Subsection:MachineParentheses}.
However, here we do without a distinction between the originally specified machine and its actual implementation --- we just put the implementation into the specified machine.

First we set up the data structure for the configurations of the machine:
\begin{verbatim}
  var t := new Gamma[a.Length + 1];
  var p:nat;
  var q:Q;
\end{verbatim}
Variable \verb!t! refers to the array containing the tape contents;
variables \verb!p! and \verb!q! contain the position of the head and the control state, respectively.

Now we load the start configuration.
The initial tape contents are the (transformed) input word followed by a single blank symbol.
Using method
\begin{verbatim}
method copyInputToTape(a:array<Sigma>, t:array<Gamma>)
  modifies t
  requires t.Length == a.Length + 1
  ensures (forall j :: 0 <= j < a.Length ==> t[j] == Sig2Gam(a[j]))
  ensures t[a.Length] == B
\end{verbatim}
we obtain the start configuration via
\begin{verbatim}
  copyInputToTape(a, t);
  p := 0;
  q := q0;
\end{verbatim}

Implementation and verification of method \verb!copyInputToTape! is again straightforward, and realizing the computation of the machine is again a direct translation of its state transition diagram, thus of Figure~\ref{fig:state.transition.diagram.TMSipserM2}:
\begin{verbatim}
  while q !in {q_acc, q_rej}
  {
    if q == q0 && t[p] == B {
      p := p + 1; q := q_rej;
    } else
    if q == q0 && t[p] == Z {
      t[p] := B; p := p + 1; q := q1;
    } else
    ...
    if q == q4 && t[p] in {Z, X} {
      if p > 0 {p := p - 1;}
    } else {
      // p := p + 1; q := q_rej;
      assert false;
    }
  }
  dec := if q == q_acc then accept else reject;
\end{verbatim}

\subsection{Verification}

We will develop our proof of correctness in five steps.
Finally at the end of each step the Dafny verification goes through.

In Step~1 we prove the absence of runtime errors.
In Step~2 we care about halving the number of zeroes on the tape.
For this we formulate and prove the `halving invariant', which is the essential ingredient for Step~3, where we formulate and prove the `power-of-two invariant'.
This invariant is in turn essential for showing the postcondition in Step~4, which concludes our proof of partial correctness.
In Step~5 we add a proof of termination and with that terminate our proof of total correctness.

Again practically:
Before showing the postcondition in Step~4 we must comment it out in the method containing the postcondition.
And before showing termination in Step~5 we must add a \verb!decreases *! clause to the loop and thus also to the method containing the loop;
this stops Dafny from attempting to prove termination.

\paragraph{Step~1: Absence of runtime errors}

Before beginning with the verification of the actual functionality of our Turing machine to decide the considered language we first ensure that our machine is free of runtime errors.
In particular, we ensure that all accesses (read or write) to the tape, \verb!t[p]!, are performed with a position \verb!p! that is in the range of the (truncated) tape.
As already mentioned, the position may exceed the tape by one square, but in this case does not access it.
So a global invariant for \verb!p! is
\begin{verbatim}
  invariant 0 <= p <= t.Length //I0
\end{verbatim}
but when accessing the tape, \verb!p! must be at most $t.\mathrm{Length} - 1 = a.\mathrm{Length}$.
This can be ensured by means of the following local invariants for all states:
\begin{verbatim}
  0 == p                 // I0.0 for q0
  1 <= p <= a.Length     // I0.1 for q1
  2 <= p <= a.Length     // I0.2 for q2
  3 <= p <= a.Length     // I0.3 for q3
  0 <= p < a.Length      // I0.4 for q4
  2 <= p == a.Length + 1 // I0.a for q_acc
  1 <= p == a.Length + 1 // I0.r for q_rej
\end{verbatim}
However, this alone does not verify.
Global invariants describing the general tape contents help:
\begin{verbatim}
  invariant t[0] == if a.Length == 0 || q != q0 then B else Z
  invariant (forall i :: 1 <= i < a.Length ==> t[i] in {Z, X})
  invariant t[a.Length] == B
\end{verbatim}

With all these invariants Dafny also verifies the \verb!assert false;! in the final else-branch of the loop body, meaning that this else-branch is unreachable.
So the only transition not explicitly handled in the loop body (the one starting from $q_0$ and reading a cross symbol) indeed can never be executed.

At this point of the verification we already know that our machine is free of runtime errors --- very comforting.

\paragraph{Step~2: Division by two and the central `halving invariant'}

Let us start the verification of the actual functionality of our machine with the division of the number of zeroes on the tape by two.
Such a division starts in state $q_1$.
If the number of zeroes currently stored on the tape is even (and at least two), the division ends up where it started: in $q_1$, and so forms a cycle --- a \emph{division cycle}.
If the number is odd (and at least three), the division ends up in state $\qrej$.

In the following we will verify that in a division cycle the (even) number of zeroes on the tape is halved, or expressed semi-formally as:
\begin{equation}\label{TMSipserM2.halving1}
\begin{split}
  &\text{number of zeroes on the tape at begin of cycle} =\\
  &\qquad 2 \cdot \text{number of zeroes on the tape at end of cycle}
\end{split}
\end{equation}
So we must compare the tape contents at one point in time with that of some other point in time.
However, nothing more than the \emph{current} tape contents are stored in the machine at any point in time.
So we take a \emph{snapshot} of the tape contents at the beginning of the cycle, store it in a ghost variable, and compare the current tape contents with this snapshot.
So we can refine our semi-formal condition (\ref{TMSipserM2.halving1}) as follows:
\begin{equation}\label{TMSipserM2.halving2}
\begin{split}
  &\text{number of zeroes on the snapshot taken at begin of cycle} =\\
  &\qquad 2 \cdot \text{number of zeroes on the tape at end of cycle}
\end{split}
\end{equation}
For storing the snapshot we declare:
\begin{verbatim}
  ghost var snapGlob:seq<Gamma>;
\end{verbatim}
and initialize the snapshot with an arbitrary value to fulfill Dafny's definite-assignment rules:
\begin{verbatim}
  snapGlob := *;
\end{verbatim}
A division cycle starts when entering state $q_1$, and this is possible initially from state $q_0$ and repeatedly from state $q_4$.
So we take a snapshot in the transition from $q_0$ to $q_1$
(\emph{after} having replaced the zero symbol with the blank symbol)
and in the transition from $q_4$ to $q_1$.
For taking a snapshot we use the command:
\begin{verbatim}
  snapGlob := t[..];
\end{verbatim}
The selfie-transition on $q_1$ does not change the tape contents, so altogether, in state $q_1$ tape contents and snapshot \emph{always} completely agree:
we obtain the following local invariant for $q_1$:
\begin{verbatim}
  (forall i :: 0 <= i < a.Length ==> snapGlob[i] == t[i]) //I1
\end{verbatim}
Now, in all of the following transitions:
$q_1 \rightarrow q_2$,
$q_2 \rightarrow q_2$,
$q_2 \rightarrow q_3$,
$q_3 \rightarrow q_3$,
$q_3 \rightarrow q_2$,
the head moves to the right, after having from time to time changed the tape symbol (from $0$ to $x$).
But only the symbol under the head is changed --- everything else, in particular the symbols to the \emph{right} of the head, remain unchanged.
Altogether, in states $q_2$ and $q_3$ tape contents and snapshot \emph{always} completely agree to the right of (and including --- the head moves \emph{after} writing) the head position.
So we obtain the following local invariant for $q_2$ and $q_3$:
\begin{verbatim}
  (forall i :: p <= i < a.Length ==> snapGlob[i] == t[i]) //I2
\end{verbatim}
Invariants $I_1$ and $I_2$ express the essence of taking the snapshots.
(By the way:
local invariant $I_1$ for state $q_1$ could be replaced by the weaker $I_2$.
Stronger invariants provide additional information and so might support a better understanding;
weaker invariants provide less superfluous information and so might support a better understanding.
It's up to you --- here and today I personally vote for the stronger $I_1$.)

Up to now, the local invariants $I_1$ and $I_2$ cannot be verified by Dafny:
indices into the snapshot might be out of range.
The following quite obvious local invariant for states $q_1$, $q_2$, and $q_3$ helps:
\begin{verbatim}
  |snapGlob| == t.Length //I3
\end{verbatim}
Next we need to formalize ``number of zeroes on the tape (snapshot)''.
Even more, for the invariants describing the ongoing division process, we need to formalize the more general ``number of zeroes in the first $i$ positions on the tape (snapshot)''.
The following two (structurally identical) ghost functions do just that:
\begin{verbatim}
ghost function numZTape(t:array<Gamma>, i:int):nat
  reads t
  requires 0 <= i <= t.Length
{
  if i == 0 then 0
  else
    if t[i-1] in {B, Z} then
      numZTape(t, i-1) + 1
    else
      numZTape(t, i-1)
}
\end{verbatim}
\begin{verbatim}
ghost function numZSnap(s:seq<Gamma>, i:int):nat
  requires 0 <= i <= |s|
{
  if i == 0 then 0
  else
    if s[i-1] in {B, Z} then
      numZSnap(s, i-1) + 1
    else
      numZSnap(s, i-1)
}
\end{verbatim}
In a division cycle the division itself is already completed when entering state $q_4$.
With all that we are now ready to formalize our semi-formal condition (\ref{TMSipserM2.halving2}) as a local invariant for state $q_4$;
we will call it the `halving invariant':
\begin{verbatim}
  numZSnap(snapGlob, a.Length) == 2 * numZTape(t, a.Length) //I4
\end{verbatim}
However, we must ensure that the precondition of function \verb!numZSnap! is fulfilled.
No problem:
invariant $I_3$ is also a local invariant of state $q_4$ and helps.

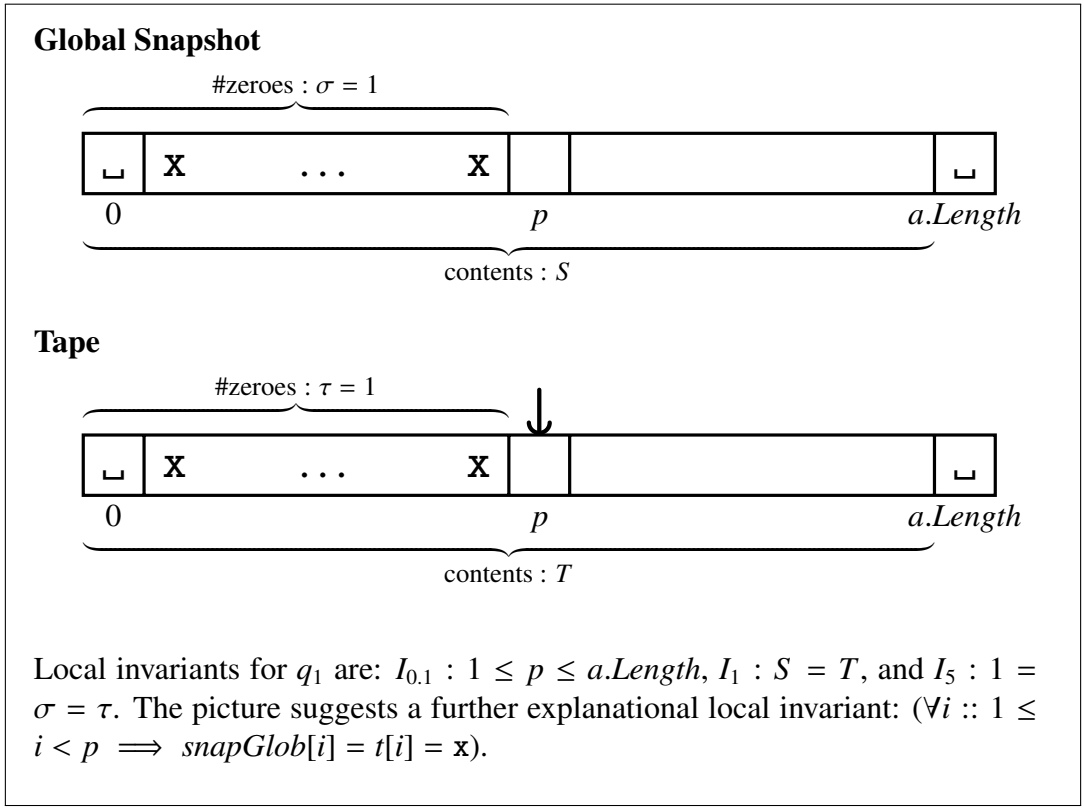
\begin{figure}[tb]
\centering
\framebox{
\begin{tikzpicture}[scale=\scalefactor,every node/.style={anchor=base}]

  \path[step=1cm,gray,very thin] (-1,-5) grid +(17,13);

  \draw (-1,2.5+\snaptapedist) node[anchor=west]
    {\textbf{Global Snapshot}};

  \draw[very thick] (0,0+\snaptapedist) rectangle +(15,1)
                    ++(1,0) -- +(0,1)
                    ++(6,0) -- +(0,1)
                    ++(1,0) -- +(0,1)
                    ++(6,0) -- +(0,1);

  \draw (0.5,0.3+\snaptapedist)
                 node {\contentssize\texttt{\sblank}}
        +(1,0)   node {\contentssize$\texttt{x}$}
        +(3.5,0) node {\contentssize{\ldots}} 
        +(6,0)   node {\contentssize$\texttt{x}$}
        +(14,0)  node {\contentssize\texttt{\sblank}};

  \draw (0.5,-0.5+\snaptapedist)
                node {\indexsize$0$}
        +(7,0)  node {\indexsize$p$}
        +(14,0) node {\indexsize$\mathit{a.Length}$};

  \draw let \n{bracewidth}={\scalefactor*7cm} in
    (3.5,1.25+\snaptapedist)
      node {$\overbrace{\makebox[\n{bracewidth}]{}}^
            {\text{\bracedescsize $\#\text{zeroes} : \sigma = 1$}}$};

  \draw let \n{bracewidth}={\scalefactor*14cm} in
    (7,-0.75+\snaptapedist)
       node {$\underbrace{\makebox[\n{bracewidth}]{}}_
             {\text{\bracedescsize contents : $S$}}$};

  \draw (-1,2.5) node[anchor=west] {\textbf{Tape}};

  \draw[very thick] (0,0) rectangle +(15,1)
                    ++(1,0) -- +(0,1)
                    ++(6,0) -- +(0,1) coordinate(p)
                    ++(1,0) -- +(0,1)
                    ++(6,0) -- +(0,1);

  \draw[ultra thick,{Arc Barb[round]}-,line cap=round]
    (p) ++(0.5,0) -- +(0,0.75);

  \draw (0.5,0.3) node {\contentssize\texttt{\sblank}}
        +(1,0)    node {\contentssize$\texttt{x}$}
        +(3.5,0)  node {\contentssize{\ldots}} 
        +(6,0)    node {\contentssize$\texttt{x}$}
        +(14,0)   node {\contentssize\texttt{\sblank}};

  \draw (0.5,-0.5) node {\indexsize$0$}
        +(7,0)     node {\indexsize$p$}
        +(14,0)    node {\indexsize$\mathit{a.Length}$};

  \draw let \n{bracewidth}={\scalefactor*7cm} in
    (3.5,1.25) node {$\overbrace{\makebox[\n{bracewidth}]{}}^
                     {\text{\bracedescsize $\#\text{zeroes} : \tau = 1$}}$};

  \draw let \n{bracewidth}={\scalefactor*14cm} in
    (7,-0.75) node {$\underbrace{\makebox[\n{bracewidth}]{}}_
                    {\text{\bracedescsize contents : $T$}}$};

  \draw let \n{minipagewidth}={\scalefactor*16.5cm} in
    (-1,2.5-\snaptapedist) node[anchor=north west] {
      \begin{minipage}{\n{minipagewidth}}
        Local invariants for $q_1$ are:
        $I_{0.1} : 1 \leq p \leq \mathit{a.Length}$,
        $I_1     : S = T$, and
        $I_5     : 1 = \sigma = \tau$.
        The picture suggests a further explanational local invariant:
        $(\forall i :: 1 \leq i < p \implies
          \mathit{snapGlob}[i] = t[i] = \texttt{x})$.
      \end{minipage}};


\end{tikzpicture}
} 
\caption{Division by two --- the local invariants for $q_1$.
  \label{fig:division.invars.q1}}
\end{figure}

\begin{figure}[tb]
\centering
\framebox{
\begin{tikzpicture}[scale=\scalefactor,every node/.style={anchor=base}]

  \path[step=1cm,gray,very thin] (-1,-4) grid +(17,12);

  \draw (-1,2.5+\snaptapedist) node[anchor=west]
    {\textbf{Global Snapshot}};

  \draw[very thick] (0,0+\snaptapedist) rectangle +(15,1)
                    ++(1,0) -- +(0,1)
                    ++(6,0) -- +(0,1)
                    ++(1,0) -- +(0,1)
                    ++(6,0) -- +(0,1);

  \draw (0.5,0.3+\snaptapedist)
                 node {\contentssize\texttt{\sblank}}
        +(14,0)  node {\contentssize\texttt{\sblank}};

  \draw (0.5,-0.5+\snaptapedist)
                node {\indexsize$0$}
        +(7,0)  node {\indexsize$p$}
        +(14,0) node {\indexsize$\mathit{a.Length}$};


  \draw let \n{bracewidth}={\scalefactor*14cm} in
    (7,-0.75+\snaptapedist)
       node {$\underbrace{\makebox[\n{bracewidth}]{}}_
             {\text{\bracedescsize $\#\text{zeroes} : \sigma$}}$};

  \draw (-1,2.5) node[anchor=west] {\textbf{Tape}};

  \draw[very thick] (0,0) rectangle +(15,1)
                    ++(1,0) -- +(0,1)
                    ++(6,0) -- +(0,1) coordinate(p)
                    ++(1,0) -- +(0,1)
                    ++(6,0) -- +(0,1);

  \draw[ultra thick,{Arc Barb[round]}-,line cap=round]
    (p) ++(0.5,0) -- +(0,0.75);

  \draw (0.5,0.3) node {\contentssize\texttt{\sblank}}
        +(14,0)   node {\contentssize\texttt{\sblank}};

  \draw (0.5,-0.5) node {\indexsize$0$}
        +(7,0)     node {\indexsize$p$}
        +(14,0)    node {\indexsize$\mathit{a.Length}$};


  \draw let \n{bracewidth}={\scalefactor*14cm} in
    (7,-0.75) node {$\underbrace{\makebox[\n{bracewidth}]{}}_
                    {\text{\bracedescsize $\#\text{zeroes} : \tau$}}$};

  \draw let \n{minipagewidth}={\scalefactor*16.5cm} in
    (-1,2.5-\snaptapedist) node[anchor=north west] {
      \begin{minipage}{\n{minipagewidth}}
        Local invariants for $q_4$ are:
        $I_{0.4} : 0 \leq p < \mathit{a.Length}$,
        and the halving invariant
        $I_4     : \sigma = 2\tau$.
      \end{minipage}};

\end{tikzpicture}
} 
\caption{Division by two --- the local invariants for $q_4$.
  \label{fig:division.invars.q4}}
\end{figure}
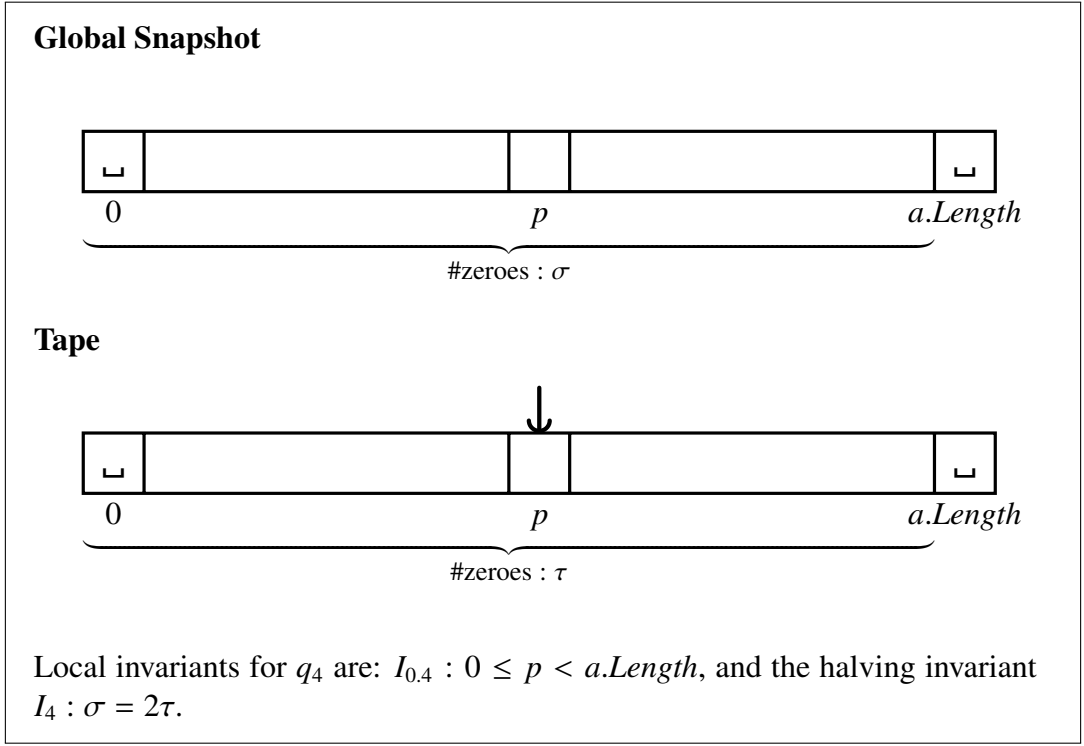

Up to now we merely \emph{suppose} that $I_4$ is an invariant.
Our next task is to show that this is indeed the case.
The division is performed by crossing off each second zero on the tape when starting from $q_1$ and then repeatedly going through $q_2$ and $q_3$.
The snapshot, on the other hand, remains unmodified.
We easily find the following local invariants relating the number of zeroes on the snapshot with the current number of zeroes on the tape between the beginning of tape or snapshot, respectively, and the current position of the head (excluding):
\begin{itemize}
\item for state $q_1$:
the number of zeroes on snapshot and tape equals one (and is thus odd) --- it is just the zero represented by the blank symbol at the beginning of the tape:
\begin{verbatim}
  1 == numZSnap(snapGlob, p) == numZTape(t, p) //I5
\end{verbatim}
\item for state $q_2$:
the number of zeroes on the snapshot is twice the number of zeroes on the tape (and is thus even):
\begin{verbatim}
  numZSnap(snapGlob, p) == 2 * numZTape(t, p) //I6
\end{verbatim}
\item for state $q_3$:
the number of zeroes on the snapshot is one less than twice the number of zeroes on the tape (and is thus odd):
\begin{verbatim}
  numZSnap(snapGlob, p) == 2 * numZTape(t, p) - 1 //I7
\end{verbatim}
\end{itemize}
(Invariant $I_3$ again ensures that the precondition of function \verb!numZSnap! is fulfilled.)
Note that whereas invariants $I_1$ and $I_2$ concern the parts of tape and snapshot to the \emph{right} of (and including) the head, invariants $I_5$, $I_6$, and $I_7$ concern the parts to the \emph{left} of (and excluding) it --- see Figures~\ref{fig:division.invars.q1}, \ref{fig:division.invars.q2.q3}, and \ref{fig:division.invars.q4}.

\begin{figure}[tb]
\centering
\framebox{
\begin{tikzpicture}[scale=\scalefactor,every node/.style={anchor=base}]

  \path[step=1cm,gray,very thin] (-1,-5) grid +(17,13);

  \draw (-1,2.5+\snaptapedist) node[anchor=west]
    {\textbf{Global Snapshot}};

  \draw[very thick] (0,0+\snaptapedist) rectangle +(15,1)
                    ++(1,0) -- +(0,1)
                    ++(6,0) -- +(0,1)
                    ++(1,0) -- +(0,1)
                    ++(6,0) -- +(0,1);

  \draw (0.5,0.3+\snaptapedist)
                 node {\contentssize\texttt{\sblank}}
        +(14,0)  node {\contentssize\texttt{\sblank}};

  \draw (0.5,-0.5+\snaptapedist)
                node {\indexsize$0$}
        +(7,0)  node {\indexsize$p$}
        +(14,0) node {\indexsize$\mathit{a.Length}$};


  \draw let \n{bracewidth}={\scalefactor*7cm} in
    (3.5,-0.75+\snaptapedist)
       node {$\underbrace{\makebox[\n{bracewidth}]{}}_
             {\text{\bracedescsize $\#\text{zeroes} : \sigma$}}$};

  \draw let \n{bracewidth}={\scalefactor*7cm} in
    (10.5,-0.75+\snaptapedist)
       node {$\underbrace{\makebox[\n{bracewidth}]{}}_
             {\text{\bracedescsize contents : $S$}}$};

  \draw (-1,2.5) node[anchor=west] {\textbf{Tape}};

  \draw[very thick] (0,0) rectangle +(15,1)
                    ++(1,0) -- +(0,1)
                    ++(6,0) -- +(0,1) coordinate(p)
                    ++(1,0) -- +(0,1)
                    ++(6,0) -- +(0,1);

  \draw[ultra thick,{Arc Barb[round]}-,line cap=round]
    (p) ++(0.5,0) -- +(0,0.75);

  \draw (0.5,0.3) node {\contentssize\texttt{\sblank}}
        +(14,0)   node {\contentssize\texttt{\sblank}};

  \draw (0.5,-0.5) node {\indexsize$0$}
        +(7,0)     node {\indexsize$p$}
        +(14,0)    node {\indexsize$\mathit{a.Length}$};


  \draw let \n{bracewidth}={\scalefactor*7cm} in
    (3.5,-0.75) node {$\underbrace{\makebox[\n{bracewidth}]{}}_
                    {\text{\bracedescsize $\#\text{zeroes} : \tau$}}$};

  \draw let \n{bracewidth}={\scalefactor*7cm} in
    (10.5,-0.75) node {$\underbrace{\makebox[\n{bracewidth}]{}}_
                    {\text{\bracedescsize contents : $T$}}$};

  \draw let \n{minipagewidth}={\scalefactor*16.5cm} in
    (-1,2.5-\snaptapedist) node[anchor=north west] {
      \begin{minipage}{\n{minipagewidth}}
        Local invariants for $q_2$ are:
        $I_{0.2} : 2 \leq p \leq \mathit{a.Length}$,
        $I_6     : \sigma = 2\tau$, and
        $I_2     : S = T$.

        Local invariants for $q_3$ are:
        $I_{0.3} : 3 \leq p \leq \mathit{a.Length}$,
        $I_7     : \sigma = 2\tau-1$, and
        $I_2     : S = T$.
      \end{minipage}};

\end{tikzpicture}
} 
\caption{Division by two --- the local invariants for $q_2$ and $q_3$.
  \label{fig:division.invars.q2.q3}}
\end{figure}
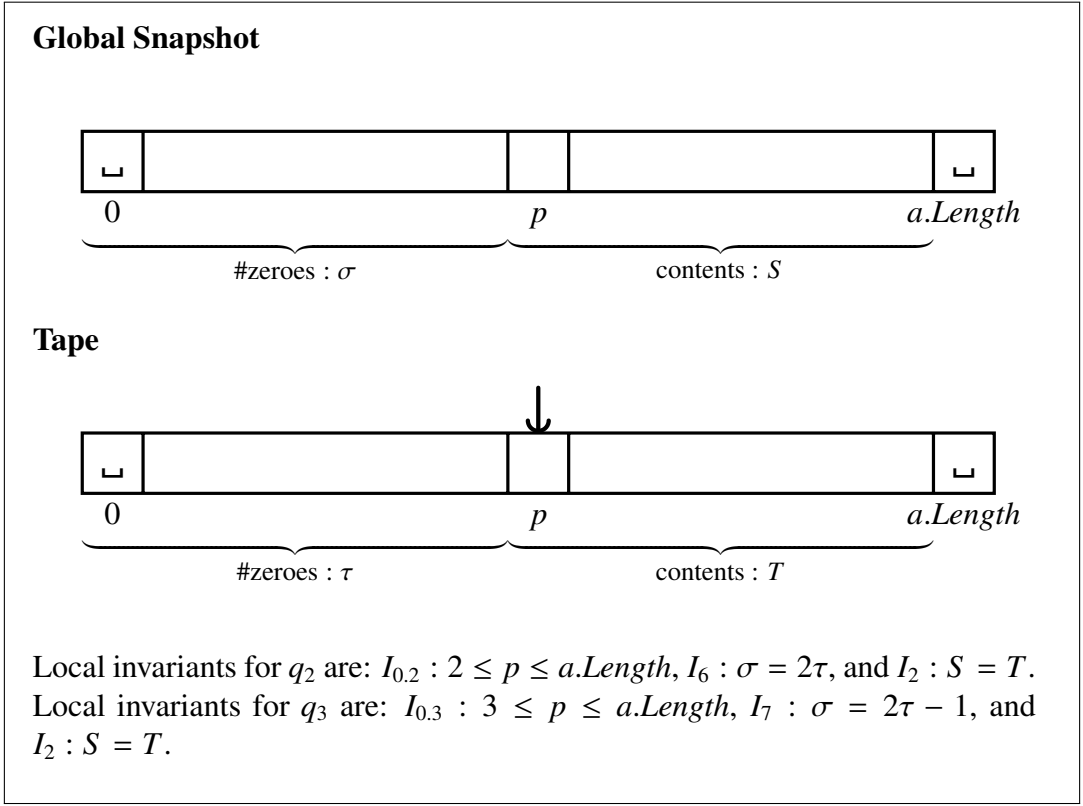

Now everything is fine --- alas, Dafny does not believe us!
The problem is this:
In the transitions $q_1 \rightarrow q_2$ and $q_3 \rightarrow q_2$ the tape contents are modified, and the tape is represented by an array.
However, a verifier considers an array not as a set of individual subscripted variables, but as a \emph{single} (array) variable, encompassing all subscripted variables.
Correspondingly, an assignment to any of the subscripted variables changes the value of the array variable as a whole.
(This is similar to, say, an integer variable to which we apply an operation that changes just a single bit at a given position --- nevertheless, the value of the variable as a whole has been changed.)

So we must convince Dafny that the considered modifications have no influence on conditions relevant to our proof.
To be more concrete:
the tape contents are modified at the head position (a zero symbol is replaced by a cross symbol) --- all other positions remain unmodified, in particular the positions \emph{before} (and excluding) the head.
Consequently, the number of zeroes on the tape before (and excluding) the head does not change.
This condition is clearly required to show maintenance of our invariants, but is not proven automatically by Dafny.

So let us embrace the assignment command `\verb!t[p] := X;!' in the two relevant transitions with a recording of the number of zeroes on the tape before (and excluding) the head position and an assertion that the number of zeroes on the tape in this range still equals the recorded number:
\begin{verbatim}
  ghost var check := numZTape(t, p); //explanational
  t[p] := X;
  assert check == numZTape(t, p); //explanational
  //assume {:axiom} check == numZTape(t, p); //explanational
\end{verbatim}
Now we should enable Dafny to prove this assertion;
and in fact, when replacing the assertion with the corresponding assumption, verification goes through --- which shows that we are on the right track.
(Program text marked as `explanational' is added to simplify understanding;
it can but need not be proved and finally must not be assumed by Dafny.)

We introduce the following quite obvious lemma:
\begin{verbatim}
lemma numZeroesLemma(t:array<Gamma>, s:seq<Gamma>, i:int)
  requires t.Length == |s|
  requires 0 <= i <= t.Length
  requires (forall j :: 0 <= j < i ==> t[j] == s[j])
  ensures numZTape(t, i) == numZSnap(s, i)
\end{verbatim}
It states that, provided tape and snapshot have the same symbols at the same first $i$ positions, the numbers of zeroes in the first $i$ positions of tape and snapshot agree.
Dafny proves this lemma automatically.

Now we embrace the assignment command (in the two relevant transitions) with a recording of a local snapshot and two applications of our lemma:
\begin{verbatim}
  ghost var check := numZTape(t, p); //explanational
  ghost var snapLocal := t[..];
  numZeroesLemma(t, snapLocal, p);
  t[p] := X;
  numZeroesLemma(t, snapLocal, p);
  assert check == numZTape(t, p); //explanational
\end{verbatim}

The first application of the lemma appears immediately after taking the snapshot.
So the precondition of the lemma is obviously satisfied, and the lemma tells us:
\begin{equation}\label{TMSipserM2.numZeroesUnchanged1}
  \mathit{numZTape}(t, p) = \mathit{numZSnap}(\mathit{snapLocal}, p)
\end{equation}
Let $t'$ denote the contents of the tape immediately after its modification.
Then the second application of the lemma could be written as
\begin{equation*}
  \mathit{numZeroesLemma}(t', \mathit{snapLocal}, p)
\end{equation*}
Since the assignment to the tape only affects position $p$, the precondition of the lemma in this second application is also satisfied, and the lemma now tells us:
\begin{equation}\label{TMSipserM2.numZeroesUnchanged2}
  \mathit{numZTape}(t', p) = \mathit{numZSnap}(\mathit{snapLocal}, p)
\end{equation}
Both equations (\ref{TMSipserM2.numZeroesUnchanged1}) and (\ref{TMSipserM2.numZeroesUnchanged2}) together yield
\begin{equation*}
  \mathit{numZTape}(t', p) = \mathit{numZTape}(t, p)
\end{equation*}
So the number of zeroes in the first $p$ positions on the tape indeed remain unchanged.
A sigh of relief seems appropriate --- now Dafny believes us!

\paragraph{Step~3: The central `power-of-two invariant'}

The central invariant of our machine states that the length of the input word is a power of two if and only if the number of zeroes on the tape before the first and after each division cycle is a power of two.
But ---as discussed in the previous step--- before entering the first (transition $q_0 \rightarrow q_1$) and after any (transition $q_4 \rightarrow q_1$) division cycle a snapshot of the current tape contents is taken.
These snapshots are then also available in states $q_2$, $q_3$, and $q_4$.
Altogether, we obtain the following local invariant for states $q_1$ up to $q_4$;
we will call it the `power-of-two invariant':
\begin{verbatim}
  isPowerOf2(numZSnap(snapGlob, a.Length)) <==>
    isPowerOf2(a.Length) //I8
\end{verbatim}

\subparagraph{Invariant holds on entry to the division cycle}
The invariant holds on entry (that is, when entering $q_1$ from $q_0$) simply because the number of zeroes on the snapshot equals the length of the input word:
\begin{equation}\label{TMSipserM2.invarOnEntry}
  \mathit{numZSnap}(\mathit{snapLocal}, a.\mathit{Length}) =
    a.\mathit{Length}
\end{equation}
Let us convince Dafny!

Immediately after copying the input word to the tape, according to the postcondition of method \verb!copyInputToTape!, the tape contains the (transformed) input word on its first $a.\mathit{Length}$ squares.
However, the \emph{only} symbol of the input alphabet is the zero symbol;
consequently, the initial contents of the first $a.\mathit{Length}$ squares of the tape are (transformed) zero symbols.
Dafny can prove the following explanational assertion without further help:
\begin{verbatim}
  copyInputToTape(a, t);
  assert (forall i :: 0 <= i < a.Length ==> t[i] == Z);
    //explanational
\end{verbatim}

We introduce the following obvious lemma:
\begin{verbatim}
lemma onlyZeroesLemma(t:array<Gamma>, i:int)
  requires 0 <= i <= t.Length
  requires (forall j :: 0 <= j < i ==> t[j] in {B, Z})
  ensures numZTape(t, i) == i
\end{verbatim}
It states that, provided the tape contains only zeroes (encoded as either blank symbols or zero symbols) on its first $i$ squares, the number of zeroes on the first $i$ squares of the tape equals $i$.
Dafny proves this lemma automatically.

Now we embrace taking the snapshot in the transition from $q_0$ to $q_1$ with an application of this lemma, an explanational assertion, and an application of the already known \verb!numZeroesLemma!:
\begin{verbatim}
  assert p == 0; //explanational
  t[p] := B;
  p := p + 1; q := q1;
  onlyZeroesLemma(t, a.Length);
  assert numZTape(t, a.Length) == a.Length; //explanational
  snapGlob := t[..];
  numZeroesLemma(t, snapGlob, a.Length);
\end{verbatim}

The precondition of the \verb!onlyZeroesLemma! is fulfilled because of the initialization of the tape contents just discussed and because the precondition properly takes the replacement of the first zero symbol by a blank symbol into account.

But wait!
The initialization of the tape contents is done outside of the loop, but the application of the \verb!onlyZeroesLemma! inside.
However, execution of a loop body in general destroys information from the outside (or from a previous loop iteration).
So to keep track of our initialization we need a further local invariant for state $q_0$:
\begin{verbatim}
  (forall i :: 0 <= i < a.Length ==> t[i] == Z) //I9
\end{verbatim}

Now, after solving this occasional problem, the \verb!onlyZeroesLemma! tells us that the number of zeroes on the tape equals the length of the input word (as confirmed by the explanational assertion):
\begin{equation*}
  \mathit{numZTape}(t, a.\mathit{Length}) = a.\mathit{Length}
\end{equation*}
Immediately after taking the snapshot, the precondition of the \verb!numZeroesLemma! is obviously fulfilled;
the lemma tells us:
\begin{equation*}
  \mathit{numZTape}(t, a.\mathit{Length}) =
    \mathit{numZSnap}(\mathit{snapLocal}, a.\mathit{Length})
\end{equation*}
The last two equations together yield our sought result (\ref{TMSipserM2.invarOnEntry}):
the invariant holds on entry.

\subparagraph{Invariant is maintained in the division cycle}
For showing that the invariant (which ---recall--- is local to states $q_1$ up to $q_4$) is maintained, we must show that it is maintained by all transitions between these states.
However, only the transition from $q_4$ to $q_1$ is crucial for the proof;
maintenance for the other transitions is trivial since they don't have any influence on the invariant.
Maintenance for the transition from $q_4$ to $q_1$ is based on the fundamental \verb!isPowerOf2LemmaEven! in cooperation with the halving invariant $I_4$ already proved in the previous step.

We embrace taking the snapshot in the transition from $q_4$ to $q_1$ with applications of the \verb!isPowerOf2LemmaEven! and the \verb!numZeroesLemma!:
\begin{verbatim}
  isPowerOf2LemmaEven(numZSnap(snapGlob, a.Length));
  snapGlob := t[..];
  numZeroesLemma(t, snapGlob, a.Length);
\end{verbatim}

Let $s$ denote the value of ghost variable \verb!snapGlob! before taking the snapshot, and $s'$ its value after taking the snapshot.
Furthermore, let us abbreviate:
\begin{align*}
  \mathit{\#ZS}  &:= \mathit{numZSnap}(s, a.\mathit{Length})\\
  \mathit{\#ZS'} &:= \mathit{numZSnap}(s', a.\mathit{Length})\\
  \mathit{\#ZT}  &:= \mathit{numZTape}(t, a.\mathit{Length})
\end{align*}
For showing maintenance of our invariant we assume that it holds (in state $q_4$) before taking the snapshot:
\begin{equation}\label{TMSipserM2.invarAssumption}
  \mathit{isPowerOf2}(\mathit{\#ZS}) \Longleftrightarrow
    \mathit{isPowerOf2}(a.\mathit{Length})
\end{equation}
and conclude that it still holds (in state $q_1$) after taking the snapshot:
\begin{equation}\label{TMSipserM2.invarConclusion}
  \mathit{isPowerOf2}(\mathit{\#ZS'}) \Longleftrightarrow
    \mathit{isPowerOf2}(a.\mathit{Length})
\end{equation}
Let us recall the halving invariant, that is, local invariant $I_4$ of state $q_4$:
\begin{equation}\label{TMSipserM2.invarDivision}
  \mathit{\#ZS} = 2 \cdot \mathit{\#ZT}
\end{equation}
The application of the \verb!isPowerOf2LemmaEven! to $\mathit{\#ZS}$ ---because of (\ref{TMSipserM2.invarDivision}) the antecedent of the postcondition of the lemma is obviously satisfied--- tells us:
\begin{equation*}
  \mathit{isPowerOf2}(\mathit{\#ZS}) \Longleftrightarrow
    \mathit{isPowerOf2}(\mathit{\#ZS} / 2)
\end{equation*}
which again using (\ref{TMSipserM2.invarDivision}) yields:
\begin{equation*}
  \mathit{isPowerOf2}(\mathit{\#ZS}) \Longleftrightarrow
    \mathit{isPowerOf2}(\mathit{\#ZT})
\end{equation*}
which in turn using the assumed invariant (\ref{TMSipserM2.invarAssumption}) yields:
\begin{equation}\label{TMSipserM2.isPO2}
  \mathit{isPowerOf2}(\mathit{\#ZT}) \Longleftrightarrow
    \mathit{isPowerOf2}(a.\mathit{Length})
\end{equation}
Using $s'$, the application of the \verb!numZeroesLemma! could be written as
\begin{equation*}
  \mathit{numZeroesLemma}(t, s', a.\mathit{Length})
\end{equation*}
This application appears immediately after taking the snapshot, so the precondition of the lemma is obviously fulfilled, and the lemma gives us:
\begin{equation*}
  \mathit{\#ZT} = \mathit{\#ZS'}
\end{equation*}
Together with (\ref{TMSipserM2.isPO2}) this yields the sought conclusion (\ref{TMSipserM2.invarConclusion}):
the invariant is maintained.

\paragraph{Step~4: Postcondition}

To conclude the proof of partial correctness we now care about the postcondition of our machine already given at the beginning of the discussion and repeated here for easier reference:
\begin{verbatim}
  ensures dec == accept <==> isPowerOf2(a.Length)
\end{verbatim}
This is an equivalence;
we prove it by showing the following two implications:
\begin{align*}
  q = \qacc &\implies \mathit{isPowerOf2}(a.\mathit{Length})\\
  q = \qrej &\implies \lnot\mathit{isPowerOf2}(a.\mathit{Length})
\end{align*}
Upon termination, according to the loop guard, $q$ equals either $\qacc$ or $\qrej$.
Thus, obviously, if $q$ is not $\qrej$, it must be $\qacc$.
Hence, from contraposition of the second implication and using double negation we obtain:
\begin{equation*}
  \mathit{isPowerOf2}(a.\mathit{Length}) \implies
   \lnot(\lnot\mathit{isPowerOf2}(a.\mathit{Length})) \implies
    \lnot (q = \qrej) \implies
     q = \qacc
\end{equation*}
The translation from the states ($\qacc$ or $\qrej$) to the $\mathit{decision}$ ($\mathit{accept}$ or $\mathit{reject}$) performed in the very last line of our program requires no further consideration.

After termination, either the number of zeroes in the input word is zero, in which case no snapshot is taken, or a snapshot is taken and the number of zeroes on the snapshot (last taken) is odd.
Whether these numbers (in input or on snapshot) are powers of two is ``known'' by the corresponding lemmas;
so we apply them after the loop:
\begin{verbatim}
  isPowerOf2LemmaZero();
  isPowerOf2LemmaOdd(numZSnap(snapGlob, a.Length));
\end{verbatim}
We will care about the precondition of function \verb!numZSnap! in the application of the `odd-lemma' (short for the \verb!isPowerOf2LemmaOdd!) alongside the following considerations.

\subparagraph{The first implication}
Let the program terminate in $\qacc$.
The (only) immediate predecessor of $\qacc$ is $q_1$.
Already proved local invariants of $q_1$ include:
\begin{align*}
  &1 \leq p \leq a.\mathit{Length}\\
  &|\mathit{snapGlob}| = t.\mathit{Length}\\
  &1 = \mathit{numZSnap}(\mathit{snapGlob}, p) = \mathit{numZTape}(t, p)
    \quad //\ I_5\\
  &\mathit{isPowerOf2}
     (\mathit{numZSnap}(\mathit{snapGlob}, a.\mathit{Length}))
       \Longleftrightarrow \mathit{isPowerOf2}(a.\mathit{Length})
\end{align*}
An already proved local invariant of $\qacc$ is
\begin{align*}
  p = a.\mathit{Length} + 1
\end{align*}
Before incrementing $p$ and entering $\qacc$, the value of $p$ was thus $a.\mathit{Length}$.
With this specialized value for $p$, from the local invariants of $q_1$ we obtain as local invariant for $\qacc$:
\begin{verbatim}
  |snapGlob| == t.Length && //I3
  1 == numZSnap(snapGlob, a.Length) &&
  (isPowerOf2(numZSnap(snapGlob, a.Length)) <==>
    isPowerOf2(a.Length))
\end{verbatim}
So in state $\qacc$ the number of zeroes on the snapshot equals one, and therefore is, according to the \verb!isPowerOf2LemmaOdd!, a power of two.
Then, finally, according to the power-of-two invariant $I_8$, $a.\mathit{Length}$ is a power of two.
So the first implication holds.

Note that the relation between the numbers of zeroes on snapshot and tape given by invariant $I_5$ gets irrelevant here;
what matters is the mere fact that the number of zeroes on the snapshot equals one.
Concerning the precondition of function \verb!numZSnap! in the application of the odd-lemma:
it is fulfilled (again) because of invariant $I_3$.

\subparagraph{The second implication}
Let the program terminate in $\qrej$.
The (only) immediate predecessors of $\qrej$ are $q_0$ and $q_3$.
State $\qrej$ is reached from $q_0$ if $a.\mathit{Length}$ equals zero;
otherwise it is reached from $q_3$.
\begin{itemize}
\item
Case $a.\mathit{Length} = 0$.
Then according to the \verb!isPowerOf2LemmaZero!, $a.\mathit{Length}$ is not a power of two.
So the second implication holds.
(That $\qrej$ is reached from $q_0$ (rather than from some other state) is not relevant: local invariants for $q_0$ are not used;
the mere fact that $a.\mathit{Length}$ equals zero suffices.)

Concerning the precondition of function \verb!numZSnap! in the application of the odd-lemma:
First, though no snapshot has been taken in this case, it \emph{has} a value, namely the unknown value given by its initialization (which, recall, is required to fulfill the definite-assignment rules).
Without any value, the application of \verb!numZSnap! would make no sense at all.
Then, the length of any sequence is non-negative;
so with $a.\mathit{Length} = 0$ the precondition of \verb!numZSnap! is trivially fulfilled.
\item
Case $a.\mathit{Length} \neq 0$.
So $\qrej$ is reached from $q_3$.
Already proved local invariants of $q_3$ include:
\begin{align*}
  &3 \leq p \leq a.\mathit{Length}\\
  &|\mathit{snapGlob}| = t.\mathit{Length}\\
  &\mathit{numZSnap}(\mathit{snapGlob}, p) =
     2 \cdot \mathit{numTSnap}(t, p) - 1 \quad //\ I_7\\
  &\mathit{isPowerOf2}
     (\mathit{numZSnap}(\mathit{snapGlob}, a.\mathit{Length}))
       \Longleftrightarrow \mathit{isPowerOf2}(a.\mathit{Length})
\end{align*}
An already proved local invariant of $\qrej$ is
\begin{align*}
  p = a.\mathit{Length} + 1
\end{align*}
Before incrementing $p$ and entering $\qrej$, the value of $p$ was thus $a.\mathit{Length}$.
With this specialized value for $p$, from the local invariants of $q_3$ we obtain as local invariant for $\qrej$:
\begin{verbatim}
  a.Length != 0 ==>
    |snapGlob| == t.Length && //I3
    numZSnap(snapGlob, a.Length) % 2 != 0 &&
    (isPowerOf2(numZSnap(snapGlob, a.Length)) <==>
      isPowerOf2(a.Length))
\end{verbatim}
So in state $\qrej$ the number of zeroes on the snapshot is odd.
Good!
But unfortunately, this condition is not strong enough:
we must also know that the number is not one.
No problem:
We add further local invariants for states $q_2$, $q_3$, $q_4$, and $\qrej$, respectively:
\begin{verbatim}
  2 <= numZSnap(snapGlob, p)          // I10.2 for q2
  3 <= numZSnap(snapGlob, p)          // I10.3 for q3
  2 <= numZSnap(snapGlob, a.Length)   // I10.4 for q4
  a.Length != 0 ==>
    3 <= numZSnap(snapGlob, a.Length) // I10.r for q_rej
\end{verbatim}
(Again: the new local invariant $I_{10.\text{\textrm{r}}}$ for $\qrej$ is a specialization of the new local invariant $I_{10.3}$ for $q_3$ with $p = a.\mathit{Length}$.)

So in state $\qrej$ the number of zeroes on the snapshot is odd \emph{and} at least three (and hence not one) and is therefore, according to the \verb!isPowerOf2LemmaOdd!, not a power of two.
Then, finally, according to the power-of-two invariant $I_8$, $a.\mathit{Length}$ is not a power of two.
So the second implication holds.

Note that the relation between the numbers of zeroes on snapshot and tape given by invariant $I_7$ gets irrelevant here;
what matters is the mere fact that the number of zeroes on the snapshot is odd.
Concerning the precondition of function \verb!numZSnap! in the application of the odd-lemma:
it is fulfilled (again) because of invariant $I_3$.

In fact invariant $I_{10.4}$ is not yet needed here;
however, it is required for the proof of termination in the next step.
\end{itemize}
This completes our proof of partial correctness.

\paragraph{Step~5: Termination}

Seen from a high-level view, progress during computation starts with taking a first snapshot, and then continues with halving the number of zeroes on the snapshots taken on successive division cycles.
In turn, progress during any halving is done by crossing off each second zero on the tape, and then by moving the head back to the beginning of the tape.
However, seen from a low-level view, progress during computation is made simply by following transition after transition.

Both views can be united by using a lexicographic ordering (quite similar to the one used in our first example machine).
Each of the components of the ordering is used to measure the progress made by following certain transitions --- corresponding to aspects of the high level view.

Our machine uses an ordering with four components, as detailed in the following.
In particular, we will detail which transitions are measured by which components.

Some of the components are conditional expressions with a branch whose value does not matter.
For writing down this branch we declare a global ghost constant $\mathit{dummy}$ of type integer with an unspecified value:
\begin{verbatim}
  ghost const dummy:int;
\end{verbatim}

\subparagraph{Component~0 \textnormal{(measuring
$q_0 \rightarrow \qrej$,
$q_0 \rightarrow q_1$)}}
First we measure the progress made when following the transitions leaving the start state: $q_0 \rightarrow \qrej$ (we are done) or $q_0 \rightarrow q_1$ (snapshots are available from now on).
Exploiting the fact that the booleans are well-ordered in Dafny, with $\mathit{false}$ less than $\mathit{true}$, the first component can be elegantly expressed as
\begin{verbatim}
  decreases q == q0
\end{verbatim}
Following any other transitions, the value of this expression remains constant --- further progress must be measured with subsequent components.

\subparagraph{Component~1 \textnormal{(measuring
$q_4 \rightarrow q_1$)}}
Here we measure the progress made in a division cycle.
Recall:
a division cycle starts in state $q_1$, goes via $q_2$ and (in general) $q_3$, and then via $q_4$ back to $q_1$.
On the final transition $q_4 \rightarrow q_1$ a new snapshot is taken, with half the number of zeroes compared to the old snapshot.
Since a division is only performed when the number of zeroes on the old snapshot is at least two, the number of zeroes strictly decreases:
\begin{verbatim}
  decreases if q != q0 then numZSnap(snapGlob, a.Length) else dummy
\end{verbatim}
On all the other transitions of the division cycle
($q_1 \rightarrow q_1$,
 $q_1 \rightarrow q_2$,
 $q_2 \rightarrow q_2$,
 $q_2 \rightarrow q_3$,
 $q_3 \rightarrow q_3$,
 $q_3 \rightarrow q_2$,
 $q_2 \rightarrow q_4$,
 $q_4 \rightarrow q_4$)
as well as on the transitions
$q_1 \rightarrow \qacc$ and
$q_3 \rightarrow \qrej$,
no snapshots are taken, so the number of zeroes on any given snapshot remains constant on these transitions --- subsequent components are required.

Dafny verifies that the precondition of function $\mathit{numZSnap}$ is satisfied whenever the function is called, that the expression strictly decreases, and that it is bounded from below by zero (even though constant $\mathit{dummy}$ is allowed to contain a negative value).

\subparagraph{Component~2 \textnormal{(measuring
$q_1 \rightarrow \qacc$,
$q_3 \rightarrow \qrej$,
$q_2 \rightarrow q_4$)}}
Here we measure the progress when following transitions $q_1 \rightarrow \qacc$ or $q_3 \rightarrow \qrej$ (we are done) or the transition $q_2 \rightarrow q_4$ (we have finished scanning the tape from left to right and start to move the head back to the beginning of the tape):
\begin{verbatim}
  decreases if q in {q1, q2, q3} then 1 else
            if q in {q4, q_acc, q_rej} then 0 else dummy
\end{verbatim}
On the remaining not yet handled transitions
$q_1 \rightarrow q_1$,
$q_1 \rightarrow q_2$,
$q_2 \rightarrow q_2$,
$q_2 \rightarrow q_3$,
$q_3 \rightarrow q_3$,
$q_3 \rightarrow q_2$,
$q_4 \rightarrow q_4$
the value of the expression remains constant --- a further component is required.

Dafny confirms that the expression strictly decreases and is bounded from below by zero.

\subparagraph{Component~3 \textnormal{(measuring
$q_1 \rightarrow q_1$,
$q_1 \rightarrow q_2$,
$q_2 \rightarrow q_2$,
$q_2 \rightarrow q_3$,
$q_3 \rightarrow q_3$,
$q_3 \rightarrow q_2$,
$q_4 \rightarrow q_4$)}}
Finally we measure the progress when following any of the not yet handled transitions simply by moving the head, square by square, in states $q_1$, $q_2$, and $q_3$ from left to right, and in state $q_4$ from right to left:
\begin{verbatim}
  decreases if q in {q1, q2, q3} then t.Length - p else
            if q in {q4} then p else dummy
\end{verbatim}

Dafny verifies that the expression strictly decreases and that it is bounded from below by zero.

\mbox{}

Now for each individual transition we have described the kind of progress it contributes to the overall computation, and we have partitioned all transitions into classes given by the components of the lexicographic ordering.
This concludes our proof of termination, and altogether, of total correctness of our Turing machine against its specification.

\section{Related Work}
\label{Section:RelatedWork}

Surprisingly, not much work has been done on the formalization, using mechanical theorem proving technology, of Turing machines.
The works I am aware of are \cite{Asperti/Ricciotti:2012:Springer} and \cite{Asperti/Ricciotti:2015:ScienceDirect} by Asperti and Ricciotti, \cite{Forster/Kunze/Wuttke:2020:ACM} by Forster, Kunze, and Wuttke, together with \cite{Wuttke:2018} by Wuttke, and \cite{Xu/Zhang/Urban:2013:Springer} by Xu, Zhang, and Urban.
All of these works provide formalisms for the systematic construction, specification, and verification of Turing machines, with the goal to simplify these tasks, or more precisely, to make these tasks feasible in the first place for non-trivial Turing machines like the universal machine.

In contrast, the work presented here takes (two) typical simple Turing machines from introductory courses on Theoretical Computer Science as examples, machines that have not at all been developed with verification in mind.
Such machines can be seen as carved out of a single stone (technically: given by their transition functions), so let us call them \emph{monolithic} machines.
In both \cite{Asperti/Ricciotti:2012:Springer} and \cite{Asperti/Ricciotti:2015:ScienceDirect} Asperti and Ricciotti write that proving the correctness of such a monolithic machine can become a ``nightmare''.
And Forster et al.\ in \cite{Forster/Kunze/Wuttke:2020:ACM} claim that ``verifying a non-trivial machine defined in terms of states and transition functions in a proof assistant seems entirely infeasible''.
Xu et al.\ in \cite{Xu/Zhang/Urban:2013:Springer} attribute the difficulties that arise to the fact that Turing machines ``can be completely \emph{unstructured}''.
The current paper seems to confirm that verifying a monolithic machine more complex than the two machines verified here is indeed infeasible, and the two verifications can be seen as a realization of such a nightmare.

\subparagraph{Asperti and Ricciotti}

In \cite{Asperti/Ricciotti:2015:ScienceDirect} Asperti and Ricciotti develop a framework for the construction, specification, and verification of multi-tape Turing machines in the Matita Theorem Prover and apply this framework to realize and prove correct a universal machines (using three tapes).
(Paper \cite{Asperti/Ricciotti:2012:Springer} is a preliminary version of this work;
in particular it still considers mono-tape rather than multi-tape machines.
Clearly, the corresponding universal machine with a single tape is considerably more complicated than the machine with three tapes.)

The model of (multi-tape) Turing machines introduced by Asperti and Ricciotti (and then further used by Forster, Kunze, and Wuttke) is different from the model of (single-tape) Turing machines introduced by Sipser and used in this paper;
besides the in general different number of tapes the essential differences are the following:
\begin{itemize}
\item
The tapes are infinite in both directions.
\item
There is no distinguished blank symbol for marking the ends of those parts of a tape that are in current use, and trying to read a symbol from a tape does not directly return a symbol, but an \emph{optional} symbol.
So when the head is in an overflow position, that is, if it tries to read a symbol when it is positioned beyond the left or right end of the tape in current use (or when this part of the tape is completely empty) it returns $\mathit{None}$ of type $\mathit{option}$ tape alphabet rather than a blank symbol of type tape alphabet.

As appears natural, Asperti and Ricciotti use a list-zipper representing a tape (together with the position of its head), with the position of the head as pivot point.
However, rather than choosing a zipper with two items and arbitrarily defining that the head is always positioned over the leftmost symbol of the right (nonempty) list (or equally well: the rightmost symbol of the left list) they use a completely symmetric and correspondingly much more elegant representation of the tape, namely a list-zipper with \emph{three} items:
the part strictly to the left of the head, the symbol under the head, and the part strictly to the right of the head (provided the head is actually positioned over the part of the tape in current use).
\item
There is no distinguished input alphabet.
\item
Rather than using a single accept and a single reject state as halting states, any state can be declared as halting state, but without distinction between accept and reject states.
\item
In addition to the two possible moves of the head to the left ($\mathrm{L}$) and the right ($\mathrm{R}$) there is a further possible (non-)move for the head to stay put ($\mathrm{N}$).
(For the ease of practically programming \emph{multi}-tape Turing machines this additional possibility is quite essential.)
\end{itemize}

The authors develop a machinery for composing larger Turing machines from smaller ones sequentially, conditionally, and iteratively (all of the involved machines having the same tape alphabet and the same number of tapes), and define a few basic machines as starting points for their constructions (like writing a symbol at the current head position of a specified tape or comparing the symbols at the current head positions of two specified tapes).

Asperti and Ricciotti specify the semantics of a Turing machine by means of a relation between the tape contents before and the tape contents after execution (with that requiring that execution terminates) and define that a machine \emph{realizes} this relation if for any initial tape contents a terminating computation exists, and initial and final tape contents are related by the relation.
Naturally, a relation realized by a larger machine composed from smaller machines is determined by operations on the relations realized by the smaller machines --- the kind of relational operations clearly correspond to the kind of composition (sequentially, conditionally, iteratively).

Interestingly, the relation to be realized ignores the control state reached after execution --- it just takes the tape contents into account.
This omission simplifies the treatment of sequential composition;
however, since the result of a guard of a conditional or iterative machine is hold in a control state, not on a tape (a design found convenient by the authors), it complicates the treatment of conditional and iterative composition.
For their treatment, Asperti and Ricciotti introduce the concept of \emph{conditional realizability}.

Finally, the authors apply their framework to the construction and verification of a universal machine;
this universal machine simulates mono-tape Turing machines given in a special normal form.

\subparagraph{Forster, Kunze, and Wuttke}

In \cite{Forster/Kunze/Wuttke:2020:ACM} Forster, Kunze, and Wuttke, starting from the work of Asperti and Ricciotti, develop a framework for the construction, specification, and verification of multi-tape Turing machines in Coq and apply their framework to realize and prove correct a universal machine.
Also they formalize a compiler from multi-tape to mono-tape machines, and their framework makes the analysis of time and space complexity possible.

The framework provides three layers of abstraction.
As base the authors use the model of Turing machines together with their elegant representation introduced by Asperti and Ricciotti.
Then, as a first layer of abstraction they introduce a labelling function mapping the control states of a machine to some type.
In computations, different control states often lead to the same control flow --- such states are mapped to the same label, and the control flow is described in terms of labels rather than control states.

The specification of the semantics of a Turing machine is, as with Asperti and Ricciotti, a relation, but now between the tape contents before execution and the \emph{pair} of the tape contents after execution and the label of the resulting control state (provided execution terminates).
The somewhat awkward concept of conditional realizability used by Asperti and Ricciotti is no longer required.

As a second layer of abstraction the authors develop a machinery for composing larger machines from smaller ones sequentially, conditionally, and iteratively, and provide basic machines as starting point for their constructions.
(The conditional composition is given in terms of a switch operation;
sequential composition and clearly a boolean conditional operation are then introduced as special cases of this switch operation.)
Control flow in conditional and iterative machines is determined by the labels introduced in the first layer of abstraction.
Furthermore, machines with different numbers of tapes or different tape alphabets can be composed using corresponding tape-lift and alphabet-lift operations;
the tape-lift operation also makes a reordering of the tapes possible.

Basic machines are mono-tape machines for reading a symbol, for writing a symbol, and for moving the head to the left or the right, and a multi-tape no-operation machine.
For example, the machine over tape alphabet $\Sigma$ for reading a symbol is a machine labelled over $\mathit{option}\ \Sigma$ with $|\Sigma| + 1$ halting states;
each halting state is labelled with $\mathit{Some}$ the symbol read, if there is one, and with $\mathit{None}$, if the head is in an overflow position (or the tape in current use is completely empty).
Here I use the names $\mathit{Some}$ and $\mathit{None}$ for the two constructors of the option type.

As a final third layer of abstraction the authors use a multi-tape Turing machine as a register machine, where each tape functions as a register which contains a value of an encodable type;
such types are the unit type, the booleans, the natural numbers, and then options, lists, sums, and products of encodable types.
Programming on this layer amounts to writing an imperative program.

Finally, the authors apply their third layer of abstraction to the construction and verification of a universal machine (using six tapes) that can simulate a mono-tape Turing machine;
the universal machine is parameterized with the tape alphabet of the simulated machine, which is thus unrestricted.

In order to exemplarily demonstrate the difference between the approaches for the verification of systematically constructed Turing machines presented in this discussion of related work and my approach for the verification of monolothic Turing machines, I have used the second layer of abstraction in the framework of Forster, Kunze, and Wuttke to rewrite our example Turing machine Sipser $M_2$.
This machine is already close to a \emph{structured} imperative program, so a rewriting for the removal of unstructured gotos was not required.
The result is presented in Appendix~\ref{Appendix:SipsersM2asFKW};
it is a machine with $56$ states altogether --- the original machine has $7$ states.

\subparagraph{Xu, Zhang, and Urban}

The work of Xu et al.\ in \cite{Xu/Zhang/Urban:2013:Springer} is essentially a formalization in the theorem prover Isabelle/HOL of the Chapters 3 up to 8 of the classical textbook \cite{Boolos/Burgess/Jeffrey:2007} about computability and logic by Boolos, Burgess, and Jeffrey.
In these chapters, Boolos et al.\ introduce three models of computation, namely Turing machines, abacus machines, and recursive functions, and show that abacus machines can be compiled to Turing machines, that recursive functions can be compiled to abacus machines, and, closing the chain, that Turing machines can be compiled to recursive functions, thus altogether showing that these three models are computationally equivalent.
(The abacus machines are register machines with an unlimited number of registers, each of which can hold a natural number of arbitrary size;
the Turing machines have tapes infinite in both directions and use a binary tape alphabet.)
Moreover, Boolos et al.\ show that the halting function is not Turing computable, and they construct a universal Turing machine.
They do the latter by first compiling a Turing machine given by its encoding (in form of a natural number) into a \emph{universal function} ---which is recursive--- and by then compiling this universal function into a Turing machine --- which then is a universal Turing machine.

The proofs given by Boolos et al.\ are presented in the usual informal mathematical style, but ---in contrast to many other textbooks that concentrate just or mainly on the overall picture--- with a remarkably careful elaboration of most (of the many) technical details.
This enables Xu et al.\ to stay quite close with their formalization to the informal presentation given in the textbook.

For formally specifying and verifying Turing machines, Xu et al.\ set up special Hoare-triples (for total correctness) and Hoare-pairs (for non-termination), a rule of consequence (for total correctness) and two rules for sequential composition (one for total correctness and one for non-termination).
They show that the two central Turing machines used in the proof of the uncomputability of the halting function given by Boolos et al., $\mathit{dither}$ and $\mathit{copy}$, are correct with respect to their specifications, which Xu et al.\ formalize as a Hoare-triple and a Hoare-pair for $\mathit{dither}$ (which may terminate or loop) and a Hoare-triple for $\mathit{copy}$ (which always terminates).
Boolos et al.\ only provide a specification for $\mathit{copy}$ and leave the design of such a machine as an exercise to the readers.
So in order to simplify the proof for $\mathit{copy}$, Xu et al.\ solve this exercise by designing the machine as sequential composition of three component machines, $\mathit{cbegin}$, $\mathit{cloop}$, and $\mathit{cend}$.
Then they individually prove the correctness of each of these components, and the correctness of $\mathit{copy}$ by using their rule of consequence and their rule for sequential composition (for total correctness).
It is interesting to note that the individual proofs of the component machines use invariants and termination metrics that have exactly the same state-dependent structure as the invariants and termination metrics used in the current paper.

\section{Conclusions}
\label{Section:Conclusions}

I reached two major conclusions:
First, verifying a Turing machine is indeed a nightmare (Section~\ref{Section:RelatedWork});
and second, using Dafny for this task transmogrifies this nightmare into a pleasure.

I have worked out the initial development of the proofs for both Turing machines using Dafny version 2.3.0.10506 under the Mono .NET framework on a MacBook Pro (2.8 GHz Quad-Core Intel Core i7 processor, 16 GB memory), where I edited the sources in Eclipse and executed Dafny in a terminal.
Recently I have migrated my development to Dafny version 4.10.0.0 (and improved it somewhat) and now run it under Visual Studio Code (with verification set to be done on save).
In both development environments the complete verification of a single Turing machine lasts just a few seconds, where the new environment is somewhat faster.
Clearly, the development under Visual Studio Code is considerably more convenient than using a terminal with an independent editor, but even that was already fun to use.

In order to explain the functioning of a Turing machine in my lectures I used some informal explanations plus some sample executions.
This procedure is also typically applied in textbooks.
For example, the functioning of Turing machine $M_2$ formally verified in this paper and originally presented in the textbook of Sipser is explained there by giving an \emph{implementation description}, that is, according to Sipser, ``English prose to describe the way that the Turing machine moves its head and the way that it stores data on its tape'' \cite[Page~157]{Sipser:2006} plus a single sample execution.
The complete discussion of this machine (definition, explanation, sample execution) covers less than two pages in Sipser's book.
And after carefully studying these two pages the functioning of the machine becomes quite clear --- big doubts concerning its correctness don't come to mind.
On the other hand, the complete discussion of Sipser's machine in this paper (definition, explanation, formal proof, \emph{no} sample executions) covers a little over $20$ pages.
So the formalization of something quite `obvious' can get really involved.
In one sense it is fascinating to observe that our intuition enables us to `understand' something without going deeply into technical details, in another sense it might be wise to raise some distrust against our intuition --- do we really grasp the whole story?

Another point to mention is a kind of asynchrony between my own and Dafny's understanding of the subject matter --- where, of course, the word `understanding' as used here has two different meanings.
Sometimes I expected Dafny to need no further help for a verification, but it did a lot, and on other occasions my expectation was vice versa, namely that Dafny would need a lot of help, but the verification immediately went through.
Concerning the pure task of providing assurance that our program is correct with respect to its specification ---which is what we want as software engineers--- we are happy when Dafny works as autonomously as possible.
On the other hand, concerning the aspect of obtaining an inherent understanding of the subject matter ---which is what we want as pondering humans--- Dafny's automation can be somewhat counterproductive, in the sense that it does not force us to cogitate enough about our subject matter.
The `never-more invariant' in our first example machine, which Dafny proves without further help, is an example in which I felt that an additional explanation could be helpful;
for this reason I did introduce the Accumulated Invariant Lemma.
Of course, there is no point in believing that such supplementary explanations indicate what Dafny \emph{really} does.

Looking at all the technical details used in the proofs presented here it becomes clear ---even considering an extremely meticulous worker--- that these details cannot be reliably handled by a human.
In particular, it is one thing to check a proof in its final form, that is, after all lemmas, invariants, bound functions, and so on are determined, but the by far bigger effort is to \emph{develop} the proof in order to determine all these lemmas and other auxiliary aids, and to perform these checks ---showing success or error--- over and over again.
This is clearly only possible using a machine, and using a tool like Dafny.

\appendix

\section{Concerning the First Machine: Parentheses}
\label{Appendix:FirstMachine}

\subsection{The Accumulated Invariant Lemma}

In the following, $\mathsf{wp}(C, Q)$ denotes the weakest precondition of a command $C$ and a postcondition $Q$ with respect to \emph{partial} correctness.

\begin{lemmaThm}[Generalized Skip Lemma]
\label{lemma:generalized.skip.lemma}
Let $T$ be a type, $k$ a variable of type $T$, $C$ a command that does not assign to $k$, and $Q$ a predicate in which $k$ may occur as a free variable, but no others.
Then
\begin{equation*}
  \mathsf{wp}(C, Q) = Q
\end{equation*}
\Halmos
\end{lemmaThm}

\begin{remarkThm}
Predicate $Q$ as precondition in general does not prevent command $C$ from looping or abortion --- so considering partial correctness is appropriate here.
\Halmos
\end{remarkThm}

\begin{proofThm}
Command $C$ has no influence on the value of variable $k$.
So if $Q$ holds before execution of $C$, it still holds after it.
And in order for $Q$ to hold after execution, it must already hold before it.
\Halmos
\end{proofThm}

\begin{lemmaThm}[Accumulated Invariant Lemma]
Given a predicate $I : \mathbb{N} \tfuntype \mathbb{B}$.
Based on this, define predicate $I_\forall : \mathbb{N} \tfuntype \mathbb{B}$ with
\begin{align*}
  I_\forall(k) = (\forall j \mid 0 \leq j \leq k \implies I(j))
\end{align*}
Given the command
\begin{align*}
  k := 0\ ;\ \mathsf{while}\ B\ \mathsf{do}\ C\ ;\ k := k + 1\ \mathsf{endwhile}
\end{align*}
where $k$ is a variable of type $\mathbb{N}$, $B$ a boolean expression, and $C$ a command that does not assign to $k$.

Let $I(k)$ be an invariant of the loop (that is, $I(k)$ is maintained by the loop body under assumption of the loop guard) and let $I(0)$ be true (that is, $I(k)$ holds on entry to the loop).
Formally:
\begin{equation*}
  I(0) \quad \text{and} \quad
  \models B \land I(k) \implies \mathsf{wp}(C\ ;\ k := k + 1, I(k))
\end{equation*}
and simplified (composition, assignment, Lemma~\ref{lemma:generalized.skip.lemma}):
\begin{equation}\label{eq:accu.invar.lemma.assumption}
  I(0) \quad \text{and} \quad
  \models B \land I(k) \implies I(k+1)
\end{equation}
Then $I_\forall(k)$ also is an invariant of the loop and $I_\forall(0)$ also is true.
Formally and simplified:
\begin{equation}\label{eq:accu.invar.lemma.conclusion}
  I_\forall(0) \quad \text{and} \quad
  \models B \land I_\forall(k) \implies I_\forall(k+1)
\end{equation}
\Halmos
\end{lemmaThm}

\begin{remarkThm}
Since $k$ is initialized with zero and incremented in the loop body, but not modified by $C$, its value is always at least zero.
We implicitly use this fact in what follows (and note that type $\mathbb{N}$ for $k$ is sensible).
\Halmos
\end{remarkThm}

\begin{remarkThm}
Notation $\models P$ says that predicate $P$ is \emph{valid}, that is, true for all possible values of all free variables occurring in $P$.
\Halmos
\end{remarkThm}

\begin{proofThm}
Concerning $I_\forall(k)$ on entry:
\begin{calculation}[\equiv]
  I_\forall(0)
\step{Def.\ of $I_\forall$ with $k = 0$}
  (\forall j \mid 0 \leq j \leq 0 \implies I(j))
\step{Arith.}
  (\forall j \mid j = 0 \implies I(j))
\step{One-point rule (GS 8.14)}
  I(0)
\end{calculation}
So if $I(0)$ holds, $I_\forall(0)$ also holds.

Concerning maintenance of $I_\forall(k)$:
We have to show that the right part of (\ref{eq:accu.invar.lemma.conclusion}) holds under the assumption that the right part of (\ref{eq:accu.invar.lemma.assumption}) holds.
We calculate:
\begin{calculation}[\equiv]
  I_\forall(k+1)
\step{Def.\ of $I_\forall$}
  (\forall j \mid 0 \leq j \leq k+1 \implies I(j))
\step{Split off term (GS 8.23)}
  (\forall j \mid 0 \leq j \leq k \implies I(j)) \land I(k+1)
\step{Def.\ of $I_\forall$}
  I_\forall(k) \land I(k+1)
\end{calculation}
and
\begin{calculation}
  I_\forall(k)
\step[\equiv]{Def.\ of $I_\forall$}
  (\forall j \mid 0 \leq j \leq k \implies I(j))
\step[\implies]{Instantiation with $j := k$ (GS 9.13)}
  0 \leq k \leq k \implies I(k)
\step[\equiv]{$k \in \mathbb{N}$}
  I(k)
\end{calculation}
Propositional calculus gives us the following implication:
\begin{equation}\label{eq:accu.invar.lemma.proposition}
  (p \land q) \land (q \implies r) \implies (p \land r)
\end{equation}
(The antecedent is true iff $p$ and $q$ and $r$ are true, and in this case the consequent is also true.)
We calculate:
\begin{calculation}
  B \land I_\forall(k)
\step[\equiv]{Idempotency of $\land$ (GS 3.38)}
  B \land I_\forall(k) \land I_\forall(k)
\step[\implies]{Proposition~(\ref{eq:accu.invar.lemma.proposition})
                with second calculation}
  B \land I_\forall(k) \land I(k)
\step[\implies]{Proposition~(\ref{eq:accu.invar.lemma.proposition})
                with Assumption~(\ref{eq:accu.invar.lemma.assumption})}
  I_\forall(k) \land I(k+1)
\step[\equiv]{first calculation}
  I_\forall(k+1)
\end{calculation}
So if $I(k)$ is maintained by the loop body under assumption of the loop guard, $I_\forall(k)$ also is.
\Halmos
\end{proofThm}

So we are done with the proof.
However, it might be interesting to see whether we could also use Dafny to mimic this proof.
Here is an attempt to do just that:
\begin{verbatim}
type T
function iniT():T

predicate I(k:int) // assumed invariant

predicate I_forall(k:int) // concluded invariant
{
  (forall j :: 0 <= j <= k ==> I(j))
}

predicate BB(x:T, k:int) // loop guard

method C_RHS(x:T, k:int) returns (r:T) // RHS of loop body

// Under the assumption that I(k) is an invariant of the loop,
// Dafny proves that I_forall(k) also is an invariant of the loop.

method accumulatedInvariantLemma()
  decreases *
{
  var x:T := iniT();
  var k := 0;
  assume {:axiom} I(0); // assumption that I(k) holds on entry
  while BB(x, k)
    decreases *
    invariant I(k)
    invariant I_forall(k)
  {
    x := C_RHS(x, k); // loop body, which does not assign to k
    k := k + 1;
    assume {:axiom} I(k); // assumption that I(k) is maintained
  }
}
\end{verbatim}

\subsection{The Remaining Cases for Showing $\Delta_a(k) \geq 0$}

\begin{itemize}
\item $q_3$ and $q_4$: Local invariants are:
$s' = s-1, k' = k+1, s \geq 0, a[k] = \rparen$.
Thus:
\begin{calculation}[\implies]
  s' = \Delta_a(k')
\step{$s' = s-1, k' = k+1$}
  s-1 = \Delta_a(k+1)
\step{Def.\ of $\Delta_a$ with $a[k] = \rparen$}
  s-1 = \Delta_a(k) - 1
\step{Arith.}
  s = \Delta_a(k)
\step{$s \geq 0$}
  \Delta_a(k) \geq 0
\end{calculation}
\item $q_5$: Local invariants are:
$s' = s, k' = k+1, s \geq 0, a[k] = \rparen$.
Thus:
\begin{calculation}[\implies]
  s' = \Delta_a(k')
\step{$s' = s, k' = k+1$}
  s = \Delta_a(k+1)
\step{Def.\ of $\Delta_a$ with $a[k] = \rparen$}
  s = \Delta_a(k) - 1
\step{$s \geq 0$}
  \Delta_a(k) \geq 1
\step{$1 \geq 0$}
  \Delta_a(k) \geq 0
\end{calculation}
\item $\qacc$: Local invariants are:
$s' = s, k' = k, s = 0$.
Thus:
\begin{calculation}[\implies]
  s' = \Delta_a(k')
\step{$s' = s, k' = k$}
  s = \Delta_a(k)
\step{$s = 0$}
  \Delta_a(k) = 0
\step{$0 \geq 0$}
  \Delta_a(k) \geq 0
\end{calculation}
\item $\qrej$ via $q_0$: Local invariants are:
$s' = s, k' = k, s \geq 1$.
Thus:
\begin{calculation}[\implies]
  s' = \Delta_a(k')
\step{$s' = s, k' = k$}
  s = \Delta_a(k)
\step{$s \geq 1$}
  \Delta_a(k) \geq 1
\step{$1 \geq 0$}
  \Delta_a(k) \geq 0
\end{calculation}
\item $\qrej$ via $q_4$: Local invariants are:
$s' = s-1, k' = k+1, s = 0, a[k] = \rparen$.
Thus:
\begin{calculation}[\implies]
  s' = \Delta_a(k')
\step{$s' = s-1, k' = k+1$}
  s-1 = \Delta_a(k+1)
\step{Def.\ of $\Delta_a$ with $a[k] = \rparen$}
  s-1 = \Delta_a(k) - 1
\step{Arith.}
  s = \Delta_a(k)
\step{$s = 0$}
  \Delta_a(k) = 0
\step{$0 \geq 0$}
  \Delta_a(k) \geq 0
\end{calculation}
\end{itemize}

\section{Concerning the Second Machine: Sipser's $M_2$}
\label{Appendix:SecondMachine}

\subsection{Proofs of the Three Lemmas}

\begin{verbatim}
lemma isPowerOf2LemmaZero()
  ensures !isPowerOf2(0)
{
  forall k ensures power(2, k) >= 1 {
    powerOf2LemmaAtLeastOne(k);
  }
}

lemma powerOf2LemmaAtLeastOne(k:nat)
  ensures power(2, k) >= 1
{}

lemma isPowerOf2LemmaEven(n:nat)
  ensures n % 2 == 0 ==> (isPowerOf2(n) <==> isPowerOf2(n / 2))
{
  isPowerOf2LemmaEvenLR(n);
  isPowerOf2LemmaEvenRL(n);
}

lemma isPowerOf2LemmaEvenLR(n:nat)
  ensures n % 2 == 0 ==> (isPowerOf2(n) ==> isPowerOf2(n / 2))
{
  assume {:axiom} n % 2 == 0;
  assume {:axiom} isPowerOf2(n);
  assert n == 2 * (n / 2);
  assert (exists k:nat :: n == power(2, k));
  var k:nat :| n == power(2, k);
  assert k >= 1;
  assert 2 * (n / 2) == 2 * power(2, k-1);
  assert n / 2 == power(2, k-1);
  assert (exists k:nat :: n / 2 == power(2, k));
  assert isPowerOf2(n / 2);
}

lemma isPowerOf2LemmaEvenRL(n:nat)
  ensures n % 2 == 0 ==> (isPowerOf2(n) <== isPowerOf2(n / 2))
{
  assume {:axiom} n % 2 == 0;
  assume {:axiom} isPowerOf2(n / 2);
  assert n == 2 * (n / 2);
  assert (exists k:nat :: n / 2 == power(2, k));
  var k:nat :| n / 2 == power(2, k);
  assert 2 * (n / 2) == 2 * power(2, k);
  assert n == power(2, k+1);
  assert (exists k:nat :: n == power(2, k));
  assert isPowerOf2(n);
}

lemma isPowerOf2LemmaOdd(n:nat)
  ensures isPowerOf2(1)
  ensures n % 2 != 0 ==> (isPowerOf2(n) ==> n == 1)
{
  isPowerOf2LemmaOne();
  isPowerOf2LemmaOddLR(n);
}

lemma isPowerOf2LemmaOne()
  ensures isPowerOf2(1)
{
  assert 1 == power(2, 0);
  assert (exists k:nat :: 1 == power(2, k));
  assert isPowerOf2(1);
}

lemma isPowerOf2LemmaOddLR(n:nat)
  ensures n % 2 != 0 ==> (isPowerOf2(n) ==> n == 1)
{
  assume {:axiom} n % 2 != 0;
  assume {:axiom} isPowerOf2(n);
  assert n == 2 * (n / 2) + 1;
  assert (exists k:nat :: n == power(2, k));
  var k:nat :| n == power(2, k);
  assert k == 0;
  assert n == 1;
}
\end{verbatim}

\section{Sipser's $M_2$ in the Formalization with Coq}
\label{Appendix:SipsersM2asFKW}

For any type $X$ and any $x$ of type $X$, Forster, Kunze, and Wuttke use notation $\coqoptiontype{X} ::= \coqnone \mid \coqsome{x}$ for type $\mathit{option}\ X$ --- $\coqnone$ for $\mathit{None}$ and $\coqsome{x}$ for $\mathit{Some}\ x$.

As already mentioned there is no need to rewrite Sipser's $M_2$ concerning structure --- it does not use unstructured jumping.
And, as already explained, the result of trying to read a symbol from the tape in the machine model of Asperti and Ricciotti and used by Forster, Kunze, and Wuttke is an element of type $\mathit{option}$ tape alphabet --- recall that this machine model has no distinguished blank symbol.

So for Sipser's $M_2$ there is no further need to initially mark the left end of the tape with a blank symbol;
now instead of rewinding the tape in internal state $q_4$ until finding the blank symbol, we rewind the tape until it overflows the left end of the tape (the head returns $\coqnone$).
However, in this overflow position we have not yet seen any zero and thus must now enter state $q_0$ rather than $q_1$ --- when rewinding to the blank symbol we had already seen a single zero (disguised as blank symbol) and thus had to enter state $q_1$.

\begin{figure}[tb]
\centering
\framebox{
\begin{tikzpicture}[scale=0.9,->,>={Stealth[scale=1.5]},
  shorten >=0pt,auto,semithick]

  \tikzstyle{every state}=[fill=black!15,draw]

  \path[step=1cm,gray,very thin] (-3,-7) grid +(16,14);

  \node[initial,state] (q0)   at +(0,0)   {$q_0$};
  \node[state]         (q1)   at +(5,0)   {$q_1$};
  \node[state]         (q2)   at +(10,0)  {$q_2$};
  \node[state]         (q3)   at +(10,-4) {$q_3$};
  \node[state]         (q4)   at +(7.5,4) {$q_4$};
  \node[state]         (qacc) at +(5,-4)  {$\qacc$};
  \node[state]         (qrej) at +(0,-4)  {$\qrej$};

  \path
    (q0) edge node
    {$\coqsome{\texttt{0}}\!\rightarrow\!\mathrm{R}$}
      (q1)
    (q0) edge node
    {$\coqnone,\coqsome{\texttt{x}}\!\rightarrow\!\mathrm{N}$}
      (qrej)
    (q1) edge node
    {$\coqsome{\texttt{0}}\!\rightarrow\!\texttt{x},\mathrm{R}$}
      (q2)
    (q1) edge node
    {$\coqnone\!\rightarrow\!\mathrm{N}$}
      (qacc)
    (q2) edge [bend left] node
    {$\coqsome{\texttt{0}}\!\rightarrow\!\mathrm{R}$}
      (q3)
    (q2) edge node [swap]
    {$\coqnone\!\rightarrow\!\mathrm{L}$}
      (q4)
    (q3) edge [bend left] node
    {$\coqsome{\texttt{0}}\!\rightarrow\!\texttt{x},\mathrm{R}$}
      (q2)
    (q3) edge [bend left] node
    {$\coqnone\!\rightarrow\!\mathrm{N}$}
      (qrej)
    (q4) edge node [swap]
    {$\coqnone\!\rightarrow\!\mathrm{R}$}
      (q0)
    (q1) edge [loop above] node
    {$\coqsome{\texttt{x}}\!\rightarrow\!\mathrm{R}$}
      ()
    (q2) edge [loop above] node
    {$\coqsome{\texttt{x}}\!\rightarrow\!\mathrm{R}$}
      ()
    (q3) edge [loop below] node
    {$\coqsome{\texttt{x}}\!\rightarrow\!\mathrm{R}$}
      ()
    (q4) edge [loop above] node
    {$\coqsome{\texttt{0}},\coqsome{\texttt{x}}\!\rightarrow\!\mathrm{L}$}
      ();

\end{tikzpicture}
} 
\caption{Sipser's $M_2$ adjusted to the formalization in Coq.
  \label{fig:state.transition.diagram.TMSipserM2Coq}}
\end{figure}
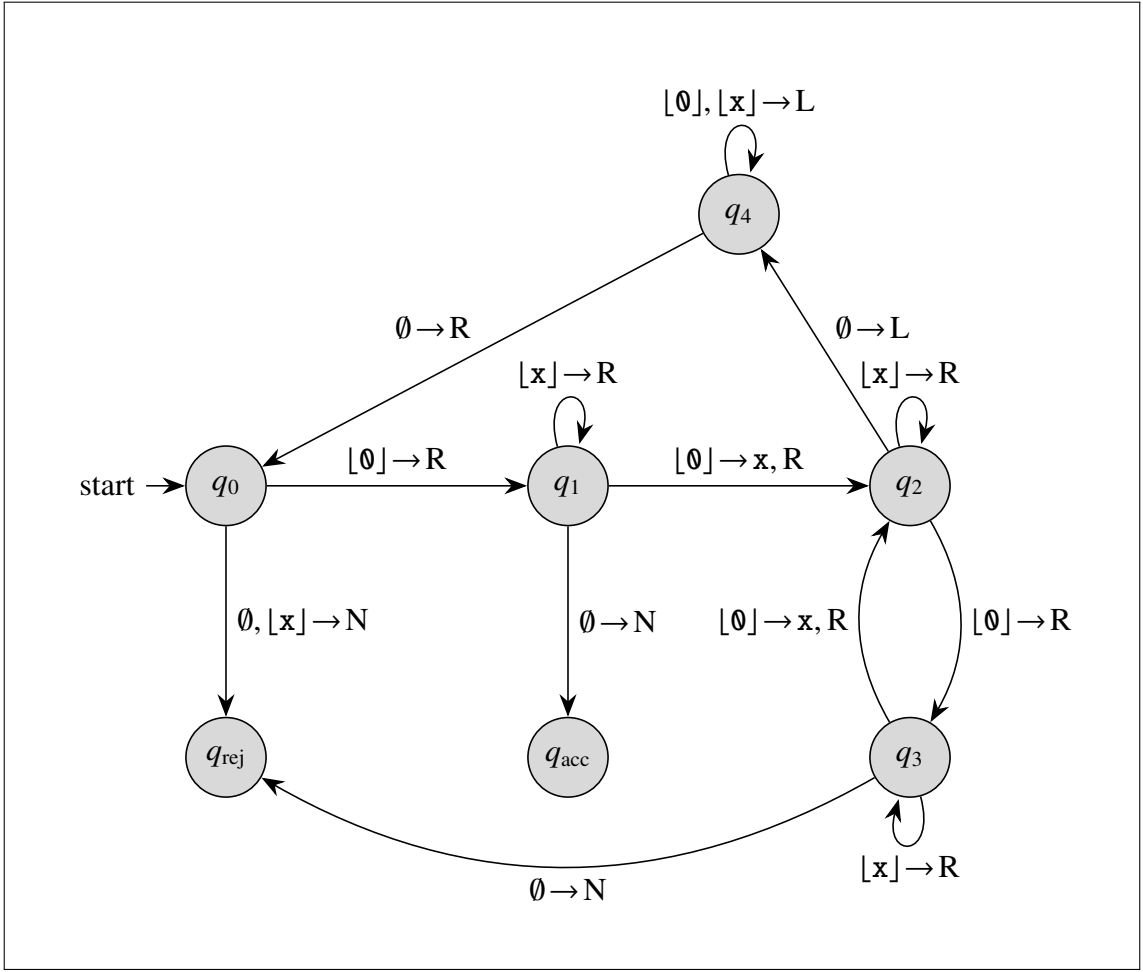

For elegance we also use the additional possibility of the new machine model for the head to stay put in all transitions leading to a halting state rather than arbitrarily moving to the right.
The resulting state transition diagram is shown in Figure~\ref{fig:state.transition.diagram.TMSipserM2Coq}.

Transforming this diagram into a machine using the second layer of abstraction of the framework of Forster, Kunze, and Wuttke is now quite straightforward.
The result is the one-tape Turing machine
$\coqTM{MSipserM2}{56} : \coqTMtypeSigmaOne(\coqlab{LabMSipserM2})$
with the labels
$\coqlab{LabMSipserM2} = \{\coqlab{Accept}, \coqlab{Reject}\}$
shown below with altogether $56$ internal states.
(I indicate the number of internal states of a machine using a superscript; this simplifies counting these numbers.)

The input alphabet of the original machine is just $\{0\}$, but in the new machine we must consider the complete tape alphabet $\{0, x\}$ as resource for the construction of input strings --- recall there is no distinguished input alphabet.
However, we are still only interested in the behaviour of the machine for input strings consisting of zeroes only.
So we can specify partial correctness (to which I want to restrict myself here) of the new machine by means of the following realization relation:
\begin{align*}
  &\coqvar{MSipserM2Rel} := \lambda\ t\ (l, t') =\\
  &\quad (\exists n \in \mathbb{N} \mid t = 0^n \land \mbox{}\\
  &\quad\quad ((\exists k \in \mathbb{N} \mid n = 2^k) \implies
                l = \coqlab{Accept}) \land \mbox{}\\
  &\quad\quad (\lnot (\exists k \in \mathbb{N} \mid n = 2^k) \implies
                l = \coqlab{Reject}))
\end{align*}
So the behaviour of the new machine remains unspecified for all input strings that contain at least one $x$.
I have not verified that the new machine indeed realizes this relation.
However, I have at least tested this realization (and termination) using a Java simulation of the machine on all input strings consisting only of zeroes whose length is between $0$ and $2^{16}$ (both bounds including).

The construction of $\coqTM{MSipserM2}{56}$ consists of altogether nine machines;
they are given in the following listing.
Each of these machines realizes the interaction between some internal states of Sipser's $M_2$ plus the actions on the outgoing transitions of these states.
For example,
machine $\coqTM{MEven}{10}$ realizes the interaction between state $q_2$ and itself plus the actions on the transitions to $q_3$, to $q_2$ (the selfie), and to $q_4$;
machine $\coqTM{MOdd}{11}$ realizes the interaction between state $q_3$ and itself plus the actions on the transitions to $q_2$, to $q_3$ (the selfie), and to $\qrej$; and
machine $\coqTM{MEvenOdd}{22}$ realizes the interaction between states $q_2$ and $q_3$ and themselves plus the actions on the transitions to $q_4$ and to $\qrej$.

Some of the machines use the relabelling operator $\coqrelabel$ defined in Wuttke~\cite{Wuttke:2018}.
This operator enables one to \emph{individually} rename labels.
In fact, operator $\coqreturnopr$ is just the special case of operator $\coqrelabel$ in which all labels are renamed with the same label.

\small

\begin{align*}
&\coqlab{LabMEven} := \{\ \coqlab{ToOdd}, \coqlab{ToRewind}\ \}\\
&\coqTM{MEven}{10} : \coqTMtypeSigmaOne(\coqlab{LabMEven}) :=
  \quad//\ q_2\\
&\quad\coqwhile\\
&\quad\quad(\coqswitch\ \coqread{4}\\
&\quad\quad\quad(\lambda\ (l : \coqoptiontype{\Sigma}).\coqmatchwith{l}\\
&\quad\quad\quad\quad\phantom{|\ }
  \coqsome{0} \Rightarrow
    \coqreturn{\coqsome{\coqlab{LabMEven}.\coqlab{ToOdd}}}{(\coqmove{R})}\\
&\quad\quad\quad\quad|\
  \coqsome{x} \Rightarrow
    \coqreturn{\coqnone}{(\coqmove{R})}\\
&\quad\quad\quad\quad|\
  \coqnone \Rightarrow
    \coqreturn{\coqsome{\coqlab{LabMEven}.\coqlab{ToRewind}}}{(\coqmove{L})}
    ))
\end{align*}

\begin{align*}
&\coqlab{LabMOdd} := \{\ \coqlab{ToEven}, \coqlab{ToReject}\ \}\\
&\coqTM{MOdd}{11} : \coqTMtypeSigmaOne(\coqlab{LabMOdd}) :=
  \quad//\ q_3\\
&\quad\coqwhile\\
&\quad\quad(\coqswitch\ \coqread{4}\\
&\quad\quad\quad(\lambda\ (l : \coqoptiontype{\Sigma}).\coqmatchwith{l}\\
&\quad\quad\quad\quad\phantom{|\ }
  \coqsome{0} \Rightarrow
    \coqreturn{\coqsome{\coqlab{LabMOdd}.\coqlab{ToEven}}}
              {(\coqwrite{x}; \coqmove{R})}\\
&\quad\quad\quad\quad|\
  \coqsome{x} \Rightarrow
    \coqreturn{\coqnone}{(\coqmove{R})}\\
&\quad\quad\quad\quad|\
  \coqnone \Rightarrow
    \coqreturn{\coqsome{\coqlab{LabMOdd}.\coqlab{ToReject}}}{\coqnop}
    ))
\end{align*}

\begin{align*}
&\coqlab{LabMEvenOdd} := \{\ \coqlab{ToRewind}, \coqlab{ToReject}\ \}\\
&\coqTM{MEvenOdd}{22} : \coqTMtypeSigmaOne(\coqlab{LabMEvenOdd}) :=
  \quad//\ q_2, q_3\\
&\quad\coqwhile\\
&\quad\quad(\coqswitch\ \coqTM{MEven}{10}\\
&\quad\quad\quad(\lambda\ (l_1 : \coqlab{LabMEven}).\coqmatchwith{l_1}\\
&\quad\quad\quad\quad\phantom{|\ }
  \coqlab{LabMEven}.\coqlab{ToOdd} \Rightarrow\\
&\quad\quad\quad\quad\quad\coqrelabel\ \coqTM{MOdd}{11}\\
&\quad\quad\quad\quad\quad\quad
  (\lambda\ (l_2 : \coqlab{LabMOdd}).\coqmatchwith{l_2}\\
&\quad\quad\quad\quad\quad\quad\quad\phantom{|\ }
  \coqlab{LabMOdd}.\coqlab{ToEven} \Rightarrow
    \coqnone\\
&\quad\quad\quad\quad\quad\quad\quad|\
  \coqlab{LabMOdd}.\coqlab{ToReject} \Rightarrow
    \coqsome{\coqlab{LabMEvenOdd}.\coqlab{ToReject}})\\
&\quad\quad\quad\quad|\
  \coqlab{LabMEven}.\coqlab{ToRewind} \Rightarrow
    \coqreturn{\coqsome{\coqlab{LabMEvenOdd}.\coqlab{ToRewind}}}{\coqnop}
    ))
\end{align*}

\begin{align*}
&\coqlab{LabMRewind} := \{\ \coqlab{ToSipserM2Body}\ \}\\
&\coqTM{MRewind}{10} : \coqTMtypeSigmaOne(\coqlab{LabMRewind}) :=
  \quad//\ q_4\\
&\quad\coqwhile\\
&\quad\quad(\coqswitch\ \coqread{4}\\
&\quad\quad\quad(\lambda\ (l : \coqoptiontype{\Sigma}).\coqmatchwith{l}\\
&\quad\quad\quad\quad\phantom{|\ }
  \coqsome{0} \Rightarrow
    \coqreturn{\coqnone}{(\coqmove{L})}\\
&\quad\quad\quad\quad|\
  \coqsome{x} \Rightarrow
    \coqreturn{\coqnone}{(\coqmove{L})}\\
&\quad\quad\quad\quad|\
  \coqnone \Rightarrow
    \coqreturn{\coqsome{\coqlab{LabMRewind}.\coqlab{ToSipserM2Body}}}
              {(\coqmove{R})}
    ))
\end{align*}

\begin{align*}
&\coqlab{LabMOdd1} := \{\ \coqlab{ToEvenOdd}, \coqlab{ToAccept}\ \}\\
&\coqTM{MOdd1}{11} : \coqTMtypeSigmaOne(\coqlab{LabMOdd1}) :=
  \quad//\ q_1\\
&\quad\coqwhile\\
&\quad\quad(\coqswitch\ \coqread{4}\\
&\quad\quad\quad(\lambda\ (l : \coqoptiontype{\Sigma}).\coqmatchwith{l}\\
&\quad\quad\quad\quad\phantom{|\ }
  \coqsome{0} \Rightarrow
    \coqreturn{\coqsome{\coqlab{LabMOdd1}.\coqlab{ToEvenOdd}}}
              {(\coqwrite{x}; \coqmove{R})}\\
&\quad\quad\quad\quad|\
  \coqsome{x} \Rightarrow
    \coqreturn{\coqnone}{(\coqmove{R})}\\
&\quad\quad\quad\quad|\
  \coqnone \Rightarrow
    \coqreturn{\coqsome{\coqlab{LabMOdd1}.\coqlab{ToAccept}}}{\coqnop}
    ))
\end{align*}

\begin{align*}
&\coqlab{LabMEven0} := \{\ \coqlab{ToOdd1}, \coqlab{ToReject}\ \}\\
&\coqTM{MEven0}{8} : \coqTMtypeSigmaOne(\coqlab{LabMEven0}) :=
  \quad//\ q_0\\
&\quad\coqswitch\ \coqread{4}\\
&\quad\quad(\lambda\ (l : \coqoptiontype{\Sigma}).\coqmatchwith{l}\\
&\quad\quad\quad\phantom{|\ }
  \coqsome{0} \Rightarrow
    \coqreturn{\coqlab{LabMEven0}.\coqlab{ToOdd1}}{(\coqmove{R})}\\
&\quad\quad\quad|\
  \coqsome{x} \Rightarrow
    \coqreturn{\coqlab{LabMEven0}.\coqlab{ToReject}}{\coqnop}\\
&\quad\quad\quad|\
  \coqnone \Rightarrow
    \coqreturn{\coqlab{LabMEven0}.\coqlab{ToReject}}{\coqnop}
    )
\end{align*}

\begin{align*}
&\coqlab{LabMEven0Odd1} :=
  \{\ \coqlab{ToEvenOdd}, \coqlab{ToAccept}, \coqlab{ToReject}\ \}\\
&\coqTM{MEven0Odd1}{20} : \coqTMtypeSigmaOne(\coqlab{LabMEven0Odd1}) :=
  \quad//\ q_0, q_1\\
&\quad\coqswitch\ \coqTM{MEven0}{8}\\
&\quad\quad(\lambda\ (l_1 : \coqlab{LabMEven0}).\coqmatchwith{l_1}\\
&\quad\quad\quad\phantom{|\ }
  \coqlab{LabMEven0}.\coqlab{ToOdd1} \Rightarrow\\
&\quad\quad\quad\quad\coqrelabel\ \coqTM{MOdd1}{11}\\
&\quad\quad\quad\quad\quad
  (\lambda\ (l_2 : \coqlab{LabMOdd1}).\coqmatchwith{l_2}\\
&\quad\quad\quad\quad\quad\quad\phantom{|\ }
  \coqlab{LabMOdd1}.\coqlab{ToEvenOdd} \Rightarrow
    \coqlab{LabMEven0Odd1}.\coqlab{ToEvenOdd}\\
&\quad\quad\quad\quad\quad\quad|\
  \coqlab{LabMOdd1}.\coqlab{ToAccept} \Rightarrow
    \coqlab{LabMEven0Odd1}.\coqlab{ToAccept})\\
&\quad\quad\quad|\
  \coqlab{LabMEven0}.\coqlab{ToReject} \Rightarrow
    \coqreturn{\coqsome{\coqlab{LabMEven0Odd1}.\coqlab{ToReject}}}{\coqnop}
    )
\end{align*}

\begin{align*}
&\coqlab{LabMEven0Odd1EvenOdd} :=
  \{\ \coqlab{ToRewind}, \coqlab{ToAccept}, \coqlab{ToReject}\ \}\\
&\coqTM{MEven0Odd1EvenOdd}{44} :
  \coqTMtypeSigmaOne(\coqlab{LabMEven0Odd1EvenOdd}) :=
  \quad//\ q_0, q_1, q_2, q_3\\
&\quad\coqswitch\ \coqTM{MEven0Odd1}{20}\\
&\quad\quad(\lambda\ (l_1 : \coqlab{LabMEven0Odd1}).\coqmatchwith{l_1}\\
&\quad\quad\quad\phantom{|\ }
  \coqlab{LabMEven0Odd1}.\coqlab{ToEvenOdd} \Rightarrow\\
&\quad\quad\quad\quad\coqrelabel\ \coqTM{MEvenOdd}{22}\\
&\quad\quad\quad\quad\quad
  (\lambda\ (l_2 : \coqlab{LabMEvenOdd}).\coqmatchwith{l_2}\\
&\quad\quad\quad\quad\quad\quad\phantom{|\ }
  \coqlab{LabMEvenOdd}.\coqlab{ToRewind} \Rightarrow
    \coqlab{LabMEven0Odd1EvenOdd}.\coqlab{ToRewind}\\
&\quad\quad\quad\quad\quad\quad|\
  \coqlab{LabMEvenOdd}.\coqlab{ToReject} \Rightarrow
    \coqlab{LabMEven0Odd1EvenOdd}.\coqlab{ToReject})\\
&\quad\quad\quad|\
  \coqlab{LabMEven0Odd1}.\coqlab{ToAccept} \Rightarrow
    \coqreturn{\coqlab{LabMEven0Odd1EvenOdd}.\coqlab{ToAccept}}{\coqnop}\\
&\quad\quad\quad|\
  \coqlab{LabMEven0Odd1}.\coqlab{ToReject} \Rightarrow
    \coqreturn{\coqlab{LabMEven0Odd1EvenOdd}.\coqlab{ToReject}}{\coqnop}
    )
\end{align*}

\begin{align*}
&\coqlab{LabMSipserM2} := \{\ \coqlab{Accept}, \coqlab{Reject}\ \}\\
&\coqTM{MSipserM2}{56} : \coqTMtypeSigmaOne(\coqlab{LabMSipserM2}) :=
  \quad//\ q_0, q_1, q_2, q_3, q_4\\
&\quad\coqwhile\\
&\quad\quad(\coqswitch\ \coqTM{MEven0Odd1EvenOdd}{44}\\
&\quad\quad\quad
  (\lambda\ (l : \coqlab{LabMEven0Odd1EvenOdd}).\coqmatchwith{l}\\
&\quad\quad\quad\quad\phantom{|\ }
  \coqlab{LabMEven0Odd1EvenOdd}.\coqlab{ToRewind} \Rightarrow
    \coqreturn{\coqnone}{\coqTM{MRewind}{10}}\\
&\quad\quad\quad\quad|\
  \coqlab{LabMEven0Odd1EvenOdd}.\coqlab{ToAccept} \Rightarrow 
    \coqreturn{\coqsome{\coqlab{LabMSipserM2}.\coqlab{Accept}}}{\coqnop}\\
&\quad\quad\quad\quad|\
  \coqlab{LabMEven0Odd1EvenOdd}.\coqlab{ToReject} \Rightarrow
    \coqreturn{\coqsome{\coqlab{LabMSipserM2}.\coqlab{Reject}}}{\coqnop}
    ))
\end{align*}

\normalsize

\bibliography{Library_DafnyfyTM}

\end{document}